\documentclass{article}
\pdfoutput=1
\usepackage{jheppub}

\usepackage{braket}
\usepackage[table]{xcolor}
\usepackage{mathrsfs}
\usepackage[english]{babel}
\usepackage{graphicx,multirow}
\usepackage{caption}
\usepackage{subcaption}
\usepackage{amssymb}
\usepackage{makecell}
\usepackage{hhline}

\usepackage[makeroom]{cancel}

\def\nn{\nonumber}

\def\pd{\partial}

\def\cD{{\cal D}}
\def\cC{{\cal C}}

\def\cI{{\cal I}}
\def\cH{{\cal H}}

\def\cO{{\cal O}}

\def\cR{{\cal R}}
\def\cT{{\cal T}}

\def\cW{{\cal W}}

\def\cS{{\cal S}}
\def\cZ{{\cal Z}}

\def\bfx{{{\bf x}}}

\def\mfm{{\mathfrak{m}}}

\def\exd{{\hbox{d}}}

\def\bea{\begin{eqnarray}}
\def\eea{\end{eqnarray}}
\def\be{\begin{equation}}
\def\ee{\end{equation}}

\def\ssH{{\scriptscriptstyle H}}
\def\ssI{{\scriptscriptstyle I}}

\def\ssQ{{\scriptscriptstyle Q}}
\def\ssR{{\scriptscriptstyle R}}
\def\ssS{{\scriptscriptstyle S}}

\def\pref#1{(\ref{#1})}
\newcommand{\roughly}[1]{\mathrel{\raise.3ex\hbox{$#1$\kern-0.85em
\lower1ex\hbox{$\sim$}}}}

\def\Tr{\mathrm{Tr}}

\def\TrB{\underset{}{\mathrm{Tr^{\,'}}}}
\def\TrAB{\underset{\rm fields}{\mathrm{Tr}}}
\def\vac{\mathrm{vac}}

\def\smath#1{\text{\scalebox{.85}{$#1$}}}
\def\sfrac#1#2{\smath{\frac{#1}{#2}}}

\numberwithin{equation}{section}
\title{Qubit Heating Near a Hotspot}
 
\date{June 2021}

\author[a,b]{G. Kaplanek,}

\author[a,b]{C.P. Burgess}

\author[c]{and R. Holman}

\affiliation[a]{Department of Physics \& Astronomy, McMaster University, 1280 Main Street West, Hamilton ON, Canada.}
\affiliation[b]{Perimeter Institute for Theoretical Physics, 31 Caroline Street North, Waterloo ON, Canada.}
\affiliation[c]{Minerva Schools at KGI,
1145 Market Street, San Francisco, CA 94103, USA.}

\abstract{Effective theories describing black hole exteriors contain many open-system features due to the large number of gapless degrees of freedom that lie beyond reach across the horizon. A simple solvable Caldeira-Leggett type model of a quantum field interacting within a small area with many unmeasured thermal degrees of freedom was recently proposed in {\tt arXiv:2106.09854} to provide a toy model of this kind of dynamics against which more complete black hole calculations might be compared. We here compute the response of a simple Unruh-DeWitt detector (or qubit) interacting with a massless quantum field $\phi$ coupled to such a hotspot. Our treatment differs from traditional treatments of Unruh-DeWitt detectors by using Open-EFT tools to reliably calculate the qubit's late-time behaviour. We use these tools to determine the efficiency with which the qubit thermalizes as a function of its proximity to the hotspot. We identify a Markovian regime in which thermalization does occur, though only for qubits closer to the hotspot than a characteristic distance scale set by the $\phi$-hotspot coupling. We compute the thermalization time, and find that it varies inversely with the $\phi$-qubit coupling strength in the standard way.}

\begin{document}
\maketitle

\section{Introduction and discussion of results}
\label{sec:intro}

The discovery of gravitational waves \cite{LIGO} adds urgency to the theoretical program to develop effective field theory (EFT) methods for physics outside a black hole, particularly in the point-particle world-line limit where the length scales of physical interest are much larger than is the black hole's horizon \cite{Goldberger:2004jt, Goldberger:2005cd, Porto:2005ac, Kol:2007bc, Kol:2007rx, Gilmore:2008gq, Porto:2008jj, Damour:2009vw, Emparan:2009at, Damour:2009wj, Levi:2015msa}.  Black holes raise new issues for EFT descriptions for several reasons. One is the practical difficulties that strong-gravity calculations raise for integrating the EFT equations of motion \cite{Allwright:2018rut, Cayuso:2017iqc, Cayuso:2020lca}. Another involves the proper EFT treatment of the dissipative degrees of freedom \cite{Goldberger:2005cd, Goldberger:2019sya} associated with the black hole's entropy. 

A challenge for developing EFTs for systems with such novel properties is the lack of theoretical benchmarks: well-understood solvable models that share some of these unusual features. Such benchmarks can be useful both as laboratories for exploring new EFTs features in these new kinds of environments, and as comparison points for calculations done with more realistic but harder-to-solve practical systems. It was with the view to providing one of these benchmarks that reference \cite{Hotspot} proposed a solvable Caldeira-Leggett style \cite{FeynmanVernon, CaldeiraLeggett} model consisting of an external massless quantum field $\phi$ interacting with many unseen gapless thermal fields in a very small spatial volume (called a `hotspot'). Some implications of this model, such as for the coherence of the state of the field $\phi$, are explored in a companion paper \cite{Hotspot:Approximate}.

The present paper explores some of the physical implications of this hotspot model by computing the response of an Unruh-DeWitt detector \cite{Unruh:1976db, DeWitt:1980hx} (or qubit) that sits at rest relative to the hotspot and separated from it by a displacement $\bfx_\ssQ$. The qubit only couples locally to the exterior field $\phi$ and so only `learns' about the hotspot through their mutual interactions with $\phi$. We use this model to study the qubit's evolution with the goal of determining whether (and how quickly) it eventually thermalizes to the hotspot temperature (as does a qubit placed outside of a spacetime horizon). 

Denoting the splitting between qubit energy levels by $\omega$ and its coupling to $\phi$ by the dimensionless coupling $\lambda_\ssQ$, we work for simplicity in a `non-degenerate' regime where the qubit's generic order-$\lambda_\ssQ^2$ field-driven energy shifts are smaller than $\omega$. It is natural to treat the field-qubit interaction strength perturbatively in this regime, with the detector's excitation rate (when prepared in its ground state) then being calculable using standard methods. 

The result for the excitation rate from the ground state is given by eqs.~\pref{11eqNZ4a} and \pref{cRIntegralResult}. As expected\footnote{This is expected because otherwise qubits at rest would continually be spontaneously excited by the vacuum, even in the absence of spacetime horizons.} this vanishes in the absence of the $\phi$-hotspot coupling $\tilde g$, and also vanishes as the hotspot temperature tends to zero. Both of these mirror properties that are also true for Unruh-DeWitt detectors in simple black-hole or cosmological backgrounds. 

The standard perturbative methods fail at times of order $1/\lambda_\ssQ^2$ --- a special case of a generic phenomenon wherein perturbation theory fails at late times --- and this means that these methods cannot directly access the time-scales relevant to hotspot thermalization, leaving the question of whether the qubit eventually reaches equilibrium beyond reach. To address this question we use open-EFT methods \cite{EFTBook, Burgess:2014eoa, Agon:2014uxa, Burgess:2015ajz, Braaten:2016sja} to resum the late-time behaviour, allowing a reliable calculation of the late-time evolution even for the $\cO(\lambda_\ssQ^{-2})$ thermalization timescales. 

Although we do not completely explore all corners of parameter space, we do identify a parameter regime --- given by \pref{parameters2} and \pref{parametersM} --- where the qubit does equilibrate at the hotspot temperature and we identify the time-scales required\footnote{Two independent and unequal time-scales arise, broadly describing both thermalization and the loss of phase coherence.} to do so --- {\it c.f.} eqs.~\pref{decaytimes} and \pref{cCIntegralResult2}. Among other things, equilibrium turns out to require both proximity to the hotspot --- with $|\bfx_\ssQ| \ll \tilde g/4\pi$ --- and sufficiently high hotspot temperatures --- $T \gg \omega$. For qubits much further from the hotspot than the hotspot's size these conditions require the response of the field $\phi$ to the hotspot to be itself understood in a regime beyond the domain of perturbation theory in $\tilde g$; a regime for which this response is nonetheless known (and given in \cite{Hotspot}) because of the solvability of the model.

The qubit behaviour we find mirrors similar thermalization behaviour found earlier for qubits in other spactimes with horizons \cite{Kaplanek:2019dqu,Kaplanek:2020iay,Kaplanek:2019vzj}, though with the important difference that unbounded redshift effects near horizons for these other spacetimes generically make thermalization more efficient near the horizon than for a hotspot. Qubit behaviour in the hotspot model is nevertheless both very rich and yet amenable to explicit calculation, and as such provides a useful test of tools that are applied in these other more complicated gravitational settings. (For other examples of qubits used to probe black-hole systems see \cite{Lin:2005uk, Hodgkinson:2012mr, Ng:2014kha, Ng:2017iqh, Emelyanov:2018woe, Jonsson:2020npo,Henderson:2019uqo,Tjoa:2020eqh,Gallock-Yoshimura:2021yok,Yu:2008zza,Hu:2011pd,Zhang:2011vsa,Hu:2012gv,Feng:2015xza,Singha:2018vaj,Chatterjee:2019kxg}.)

The remainder of this paper is organized as follows. The setup of the hotspot system is first summarized in \S\ref{sec:setup}, culminating with expressions \pref{fullW}, \pref{curlySanswer} and \pref{curlyEanswer} giving explicit formulae for the late-time $\phi$ response function in position space. For comparison, the perturbative limit of this expression is also given in eq.~\pref{Wpertfull}. This is followed in \S\ref{sec:Qubit} by the definition of the qubit and its couplings to the hotspot. The response of the qubit is also calculated in this section, both perturbatively and after resumming using open EFT techniques. This late-time resummation hinges on a late-time limit in which the qubit evolution becomes Markovian, and so considerable care is taken to justify the domain of validity of this approximation. 

\section{Hotspot properties}
\label{sec:setup}

This section briefly reviews the main features of the benchmark hotspot model proposed in \cite{Hotspot}, whose interactions with the Unruh-DeWitt detector are to be studied.  

\subsection{Hotspot definition}
\label{ssec:HotspotDef}

\begin{figure}[h]
\begin{center}
\includegraphics[width=60mm,height=60mm]{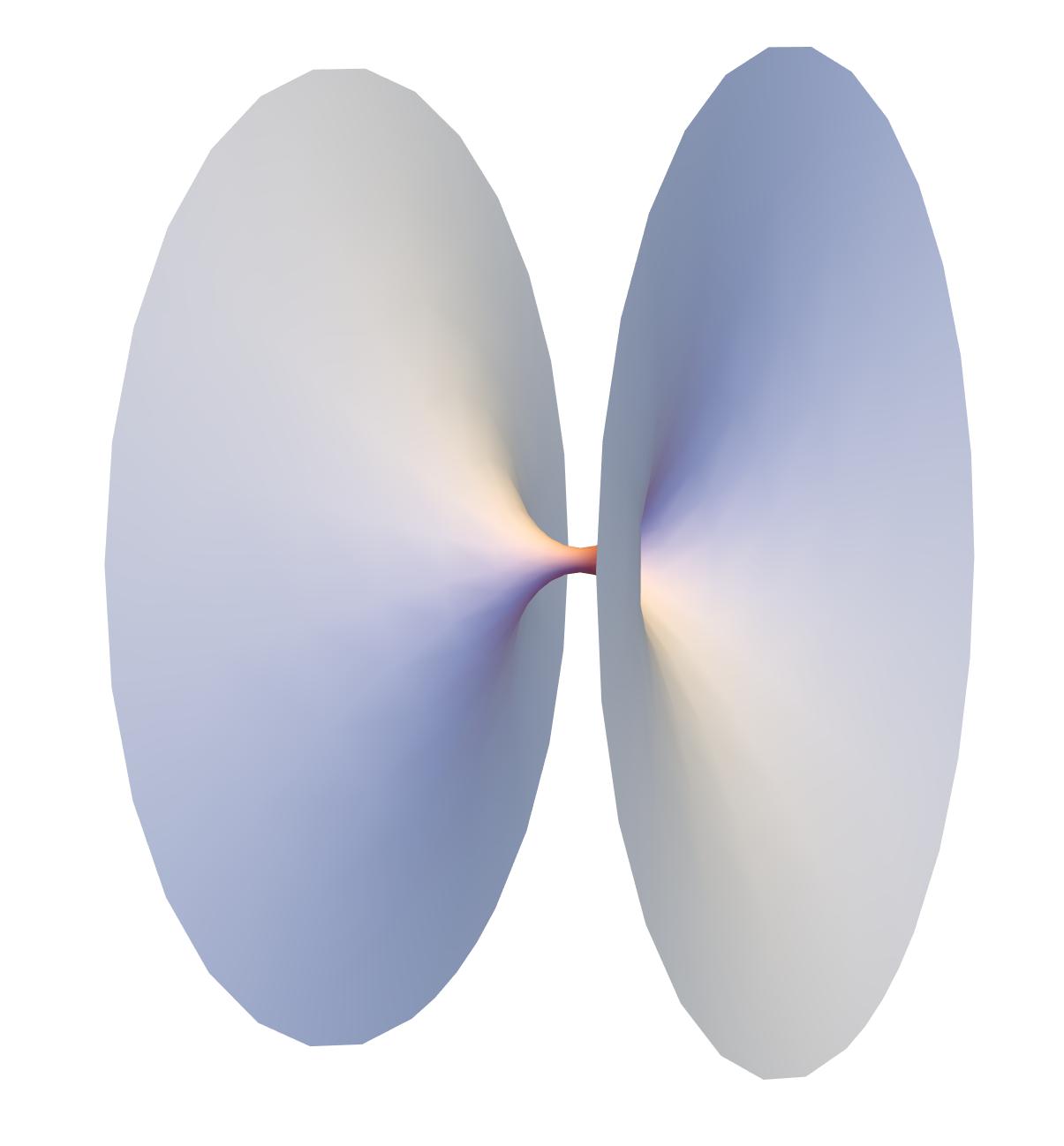} 
\caption{A cartoon of the two spatial branches, $\cR_+$ and $\cR_-$, in which the field $\phi$ and the $N$ fields $\chi^a$ repsectively live. The two types of fields only couple to one another in the localized throat region, which can be taken to be a small sphere of radius $\xi$, or effectively a point in the limit that $\xi$ is much smaller than all other scales of interest. (Figure taken from \cite{Hotspot}.)} \label{fig:FunnelFig} 
\end{center}
\end{figure}

The hotspot is taken to contain an observable sector, modelled by a single real scalar field, $\phi(x)$, that lives in a spatial region, $\cR_+$, that represents the exterior of the black hole. The degrees of freedom interior to the black hole is modelled by $N$ real massless scalar fields, $\chi^{a}$ with $a=1,\cdots,N$, that reside in a different spatial region $\cR_-$ that is disjoint from the region $\cR_+$ everywhere except for the surface of a small sphere, $\cS_\xi$, with radius $\xi$. In practice this means that both $\cR_+$ and $\cR_-$ have a small sphere excised from the origin (for all time) and the surface of this sphere is identified in the two spaces (see Figure \ref{fig:FunnelFig}). 

The fields are allowed to interact with one another locally only on $\cS_\xi$, but to keep the model solvable this interaction is limited to a bilinear mixing term. Although we neglect the external gravitational fields of the hotspot in regions $\cR_+$ and $\cR_-$ there is no reason why this could not also be included in more sophisticated versions of the model.\footnote{Without a strong gravitational field the interaction surface $\cS_\xi$ is generically not light-like and so is not a local horizon, unlike for an honest-to-God black hole.}  Our interest in this paper is in scales much larger than $\xi$ and so we further consider the idealization\footnote{The limit $\xi \to 0$ is not required for the hotspot model, allowing it also to explore the opposite regime where UV scales involve distances much smaller than $\xi$, such as for the near-horizon EFTs considered in \cite{Burgess:2018pmm, Rummel:2019ads}, motivated to systematize the treatment of both conventional \cite{Price:1986yy, Thorne:1986iy, Damour:1978cg, Parikh:1997ma, Donnay:2019jiz} or more exotic \cite{Cardoso:2016rao, Abedi:2016hgu, Holdom:2016nek, Cardoso:2017cqb, Bueno:2017hyj, Mark:2017dnq, Conklin:2017lwb, Berti:2018vdi, Zhou:2016hsh} kinds of near-horizon physics.} where the radius $\xi \to 0$, in which case $\cS_\xi$ reduces to a single point of contact between $\cR_+$ and $\cR_-$ (which we situate at the origin $\bfx = \mathbf{0}$ of both $\cR_\pm$). In this limit the couplings between $\phi$ and $\chi^a$ are captured by an effective action localized at $\bfx = 0$. 

The action has the form $S = S_+ + S_- + S_{\rm int}$ where $S_\pm$ describe the free fields $\phi$ and $\chi^a$ 
\be \label{Naction}
S_+ = - \frac{1}{2} \int_{\cR_+} \exd^{4}x \; \partial_{\mu} \phi \, \partial^{\mu} \phi \quad\hbox{and}\quad
S_- =  -\frac12 \int_{\cR_-} \exd^4x\; \delta_{ab} \, \partial_{\mu} \chi^{a} \partial^{\mu} \chi^{b}  \,,
\ee
and the lowest-dimension interaction (mixing, really) on the interaction surface is given by
\be \label{Sint3}
S_{\mathrm{int}} =   - \int_\cW   \exd t \;\left[  g_a \, \chi^{a}(t,\mathbf{0}) \,  \phi(t,\mathbf{0}) + \frac{\lambda}2 \, \phi^2(t,\mathbf{0}) \right] \,,  
\ee
where the integration is over the proper time along the hotspot world-line $\cW$, which we take to be $\bfx = 0$ in both $\cR_+$ and $\cR_-$. In fundamental units the couplings $g_a$ and $\lambda$ and have dimensions of length. 

In what follows we imagine both of these couplings turn on suddenly at $t = 0$ --- {\it i.e.} we assume $g_a(t) = \Theta(t) \, g_a$, where $\Theta(t)$ is the Heaviside step function --- but remain constant thereafter. Because our applications focus on a qubit that couples only to $\phi$, the couplings $g_a$ often appear only through the combination
\be
  \tilde g^2  := \delta^{ab} g_a g_b = N g^2 \,,
\ee
where the second equality specializes to the case where all couplings are equal (as we typically do). Because we solve for the evolution of $\phi$ exactly we need not assume that $\tilde g$ be particularly small. 

\subsection{Time evolution and Wightman function}
\label{subsec:schropic}

Ref.~\cite{Hotspot} computes the evolution of the system after the couplings $\lambda$ and $g_a$ are turned on, assuming the system's initial state at $t = 0$ is initially uncorrelated
\be \label{initialstate}
\rho_0 = \rho_+ \otimes \rho_-  \,,
\ee
with the $\phi$ sector initially in its vacuum and the $\chi^a$ fields initially in a thermal state:
\be \label{initialstate+-}
\rho_+ = \ket{\vac} \bra{\vac}  \quad\hbox{and} \quad
\rho_- = \varrho_{\beta} := \frac{e^{ - \beta \cH_{-} } }{ \cZ_\beta}  \,,
\ee
with inverse temperature $\beta = 1/T > 0$. Here $\cH_-$ denotes the Hamiltonian constructed from $S_-$ and $\cZ_\beta := \TrB[ e^{ - \beta \cH_{-}} ]$ is the thermal partition function, with the prime on the trace indicating that it is only taken over the $\chi$ sector. 

Ref.~\cite{Hotspot} computes the system response by solving explicitly the field equations 
\be \label{heis1}  
( - \partial_t^2 + \nabla^2 ) \phi _{\ssH}(t,\bfx) = \delta^{3}(\bfx) \bigg[ \lambda \phi _{\ssH}(t,\mathbf{0}) +  g_a \chi_{\ssH}^a(t,\mathbf{0}) \bigg] 
\ee
and
\be \label{heis2}
( - \partial_t^2 + \nabla^2 ) \chi^{a}_{\ssH}(t,\bfx) = \delta^{3}(\bfx) \; g_a \phi _{\ssH}(t,\mathbf{0})  \,,
\ee
within the Heisenberg picture of time evolution. This can be done very explicitly because the field equations are linear in all of the fields. To compute the response of the $\phi$ field the field $\chi^a$ is eliminated by solving \pref{heis2} for it as a function of $\phi$.

Because the elimination of $\chi^a$ involves Coulomb-like Greens functions proportional to $1/|\bfx|$ it introduces singularities into the solution for $\phi$ at $|\bfx| = 0$ that are regulated by instead evaluating at $|\bfx| = \epsilon$ for a microscopic scale $\epsilon$. Ref.~\cite{Hotspot} shows that physical predictions remain independent of $\epsilon$ once the singular regularization dependence is absorbed by replacing $\lambda \to \lambda_\ssR$ with
\be
\lambda_{\ssR} := \lambda - \frac{\tilde{g}^2}{4 \pi \epsilon} \ ,
\ee
and this is why this particular coupling is included in addition to $\tilde g$. In what follows we assume this replacement has been done, though we drop the subscript `$R$' to avoid notational clutter.

These steps allow the calculation of the $\phi$-field Wightman function, 
\be \label{CorrelationStart}
W_\beta(t,\bfx; t',\bfx') := \Tr\Bigl[ \phi _{\ssH}(t,\bfx) \phi _{\ssH}(t',\bfx') \rho_0 \Bigr] = \frac{1}{Z_\beta} \mathrm{Tr}\Bigl[ \phi _{\ssH}(t,\bfx) \phi _{\ssH}(t',\bfx') \big( \ket{\mathrm{vac}} \bra{\mathrm{vac}} \otimes  e^{ - \beta \cH_{-}} \big) \Bigr] \,.
\ee
The result computed to leading-order in $\tilde{g}^2$ and $\lambda$ turns out to be given by
\bea \label{Wpertfull}
&\ & W_\beta(t,\bfx; t',\bfx') \simeq \frac{1}{4\pi^2 \big[ - (t - t' - i \delta)^2 + |\bfx - \bfx'|^2 \big]}\nn \\ 
&\ & \qquad + \frac{\lambda}{16\pi^3} \bigg( \frac{\Theta(t-|\bfx|)}{|\bfx|} \frac{1}{(t-t'-|\bfx| - i \delta)^2 - |\bfx'|^2} + \frac{\Theta(t'-|\bfx'|)}{|\bfx'|} \frac{1}{(t-t'+|\bfx'| - i \delta)^2 - |\bfx|^2} \bigg) \nn \\
& \ & \qquad - \frac{\tilde{g}^2 \Theta(t-|\bfx|) \Theta(t' - |\bfx'|)}{64 \pi^2 \beta^2 |\bfx| |\bfx'| \sinh^2 \left[ \frac{\pi}{\beta}( t - |\bfx| - t' + |\bfx'| - i \delta ) \right]}  \\
&\ & \qquad + \frac{\tilde{g}^2}{32 \pi^4} \bigg( - \frac{\Theta(t-|\bfx|)}{|\bfx|} \frac{t-t'-|\bfx|}{\big[ (t-t'-|\bfx| - i \delta)^2 - |\bfx'|^2 \big]^2} + \frac{\Theta(t'-|\bfx'|)}{|\bfx'|} \frac{t-t'+|\bfx'|}{\big[ (t-t'+|\bfx'| - i \delta)^2 - |\bfx|^2 \big]^2} \bigg) \nn \\
&\ & \qquad + \frac{\tilde{g}^2}{64\pi^4} \bigg( \frac{\delta(t - |\bfx|)}{|\bfx| \big[ - (t' + i \delta)^2 - |\bfx'|^2 \big]} + \frac{\delta(t' - |\bfx'|)}{|\bfx'| \big[ - (t - i \delta)^2 - |\bfx|^2 \big]} \bigg) \qquad \hbox{(perturbative)}\,, \nn
\eea
where the delta functions and step functions describe the passage of the transients from the switch-on of couplings at $t = |\bfx| = 0$, followed by late-time evolution in the future light cone of the switch-on event ({\it i.e.} for $t>|\bfx|$ and $t' > |\bfx'|$). In this late-time regime the above perturbative expression becomes
\bea  \label{pertcorr}
W_\beta(t,\bfx; t',\bfx') & \simeq & \frac{1}{4\pi^2 \big[ - (t - t' - i \delta)^2 + |\bfx - \bfx'|^2 \big]} + \frac{\lambda}{16\pi^3 |\bfx| |\bfx'|} \bigg[ \frac{|\bfx| + |\bfx'|}{(t - t' - i \delta)^2 - (|\bfx + |\bfx'|)^2 } \bigg]\nn \\
& \ & \qquad - \frac{ \tilde{g}^2 }{64 \pi^2 \beta^2 |\bfx| |\bfx'| \sinh^2 \left[ \frac{\pi}{\beta}( t - |\bfx| - t' + |\bfx'| - i \delta ) \right]} \\
&\ & \qquad + \frac{\tilde{g}^2}{32 \pi^4} \bigg( - \frac{1}{|\bfx|} \frac{t-t'-|\bfx|}{\big[ (t-t'-|\bfx| - i \delta)^2 - |\bfx'|^2 \big]^2} + \frac{1}{|\bfx'|} \frac{t-t'+|\bfx'|}{\big[ (t-t'+|\bfx'| - i \delta)^2 - |\bfx|^2 \big]^2} \bigg) \nn\\
&& \qquad \qquad \qquad\qquad \qquad \qquad \qquad \qquad \qquad \qquad  \qquad\qquad \qquad \hbox{(late times, perturbative)}\nn
\eea
In these expressions $\delta \to 0^+$ is a positive infinitesimal that is taken to zero at the end of the calculation.

But the simplicity of the model allows a more general determination of the Wightman function in the future light-cone of $t= |\bfx| = 0$ that is not restricted to perturbatively small couplings. This more exact treatment gives (for $t > |\bfx|$ and $t' > |\bfx'|$)
\be \label{fullW}
W_{\beta}(t,\bfx ; t', \bfx') = \mathscr{S}(t,\bfx ; t, \bfx') + \mathscr{E}_\beta(t,\bfx ; t, \bfx') 
\ee
with the temperature-independent part given by
\bea  \label{curlySanswer} 
\mathscr{S}(t,\bfx ; t, \bfx') & = & \frac{1}{4 \pi^2 \left[ - (t - t' - i \delta)^2 + |\bfx - \bfx'|^2 \right]} \nn\\
& \ & \qquad + \frac{ 2 \epsilon^2}{\tilde{g}^2 |\bfx| |\bfx'|} \bigg[ I_{-}(t-t'+|\bfx|+|\bfx'|, c) - I_{-}(t-t'-|\bfx|+|\bfx'|, c) \\
& \ & \qquad \qquad \qquad \qquad \qquad \qquad - I_{+}( t - t' - |\bfx| + |\bfx'| , c) + I_{+}( t - t' - |\bfx| - |\bfx'| , c) \bigg] \nn \\
& \ & \qquad + \frac{\epsilon}{8 \pi^2 |\bfx| |\bfx'|} \bigg[ - \frac{1}{t - t' + |\bfx| + |\bfx'| - i \delta} + \frac{1}{t - t' - |\bfx| - |\bfx'| - i \delta } \bigg] \nn \\
& \ & \qquad \quad -  \frac{32 \pi^2 \epsilon^4 ( 1 + \frac{\lambda}{2 \pi \epsilon} )}{ \tilde{g}^4 |\bfx| |\bfx'|} \bigg[ I_{-}( t - t' -|\bfx| + |\bfx'| , c ) + I_{+}( t - t' -|\bfx| + |\bfx'| , c )  \bigg] \nn \\
& \ & \qquad \qquad - \frac{\epsilon^2}{4 \pi^2 |\bfx| |\bfx'| ( t - t' -|\bfx| + |\bfx'| - i \delta )^2} \nn
\eea
where
\be
c := \frac{16 \pi^2 \epsilon}{\tilde{g}^2} \left( 1  + \frac{\lambda}{4 \pi \epsilon} \right) \ . \label{cdef}
\ee
and the functions $I_{\mp}(\tau)$ are defined by
\be \label{IpmDef}
I_{\mp}(\tau,c) = e^{\pm c \tau} E_{1}\big( \pm c [\tau - i \delta] \big) =  e^{\pm c \tau} \int_z^\infty \exd u\; \frac{e^{-u}}{u} \,,
\ee
and the limit $\delta \to 0^{+}$ is again understood. The temperature-dependent\footnote{Notice that although $\mathscr{E}_\beta$ contains all of the dependence on temperature it does not vanish in the $T \to 0$ limit.} part is similarly given by
\be \label{curlyEanswer}
\mathscr{E}_\beta(t,\bfx ; t, \bfx') = \frac{2\epsilon^2}{\tilde{g}^2 |\bfx| |\bfx'|} \bigg[ \Psi\bigg( e^{ - \tfrac{2\pi (t - t' - |\bfx| + |\bfx'| - i \delta)}{\beta}}, \frac{c\beta}{2\pi} \bigg) + \Psi\bigg( e^{ + \tfrac{2\pi (t - t' - |\bfx| + |\bfx'| - i \delta)}{\beta}}, \frac{c\beta}{2\pi} \bigg) \bigg]  - \frac{2 \pi}{c\beta} \bigg] 
\ee
with $\Psi(z,a) := \Phi(z,1,a)$ where $\Phi(z,s,a)$ is the Lerch transcendent, defined by the series 
\be \label{PhiDef}
   \Phi(z,s,a) := \sum_{n=0}^{\infty} \frac{z^n}{(a+n)^s} 
\ee
for complex numbers in the unit disc (with $|z|<1$), and by analytic continuation elsewhere in the complex plane. A convenient integral representation for $\Phi(z,s,a)$ is given by
\be\label{PhiIntRep}
\Phi(z,s,a) \ = \ \frac{1}{\Gamma(s)} \int_0^\infty \exd x \; \frac{x^{s-1} e^{-ax} }{ 1 - z e^{-x} } \qquad \mathrm{valid\ for\ }\mathrm{Re}[s] > 0,\ \mathrm{Re}[a]>0\ \& \ z\in \mathbb{C} \setminus [1,\infty) \,.
\ee

The full correlation function $W_{\beta} = \mathscr{S} + \mathscr{E}_\beta$ obtained using \pref{curlySanswer} and \pref{curlyEanswer} reduces to the perturbative correlation function quoted in \pref{pertcorr} once linearized in $\lambda$ and $\tilde g^2$, and seeing how this works clarifies the domain of validity of perturbative methods. The expansion in $\tilde g^2$ in particular is captured by the asymptotic form for $\mathscr{S}$ in the regime $c \tau \gg 1$ as well as the expansion of $\mathscr{E}_\beta$ in the regime $c\beta \gg 1$. These can be made explicit using the asymptotic expression
\be
E_{1}(z)\ \simeq \ e^{ - z} \bigg[ \frac{1}{z} - \frac{1}{z^2} + \cO \left(z^{-3} \right)  \bigg] \qquad \qquad \mathrm{for}\ |z| \gg 1
\ee
which implies that the functions $I_{\mp}(\tau, c )$ for $| c \tau | \gg 1$ have the asymptotic expansion
\be
I_{\mp}(\tau, c) \ \simeq \ \pm \frac{1}{c(\tau - i \delta)} - \frac{1}{c^2(\tau - i \delta)^2} + \cO\left(|c\tau|^{-3}\right) \qquad \qquad \mathrm{for}\  |c \tau| \gg 1 \ .
\ee
The behaviour of $\Psi(z,a)$ for $c\beta \gg 1$ is similarly given by the following asymptotic series for the Lerch transcendent for large positive $a$:
\be
\Phi(z,s,a) \simeq \frac{a^{-s}}{1-z}  + \sum_{n=1}^{N-1} \frac{(-1)^n \Gamma(s+n)}{n!\; \Gamma(s)} \cdot  \frac{\mathrm{Li}_{-n}(z)}{a^{s+n}} + \cO \left(a^{-s-N}\right) \qquad \quad \mathrm{for\ } a \gg 1
\ee
which applies for fixed $s \in \mathbb{C}$ and fixed $z \in \mathbb{C}  \setminus [ 1, \infty )$, where $\mathrm{Li}_{-n}(z) = \left( z \partial_{z} \right)^n \frac{z}{1-z}$ are polylogarithm functions of negative-integer order. These and some other properties are explored in Appendix \ref{App:modesumfull}.

Besides verifying that the apparent $\epsilon$-dependence cancels in $W_\beta$ in the perturbative limit, the above expressions show that the perturbative limit arises as an expansion in powers of   
\be \label{PertParams}
  \frac{1}{c\tau} = \frac{\tilde g^2}{ 16 \pi^2 \epsilon \tau} \left( 1 + \frac{\lambda}{4\pi \epsilon} \right)^{-1} \ll 1 \quad \hbox{and} \quad
   \frac{1}{c\beta} = \frac{\tilde g^2 T}{ 16 \pi^2 \epsilon} \left( 1 + \frac{\lambda}{4\pi \epsilon} \right)^{-1} \ll 1 \,,
\ee
which includes low temperatures ($T$) and long times ($\tau$) compared with the UV scale $\tilde g^2/4\pi \epsilon$. As shown in \cite{Hotspot} the dependence of \pref{PertParams} on $\lambda/4\pi \epsilon$ is properly captured by renormalization-group methods in the world-line EFT for this system. 

\section{Response of an Unruh-DeWitt detector}
\label{sec:Qubit}

This section couples a simple two-level qubit (or Unruh-DeWitt detector) that moves in $\cR_+$ near the hotspot but not on the interaction surface $\cS_\xi$, coupling locally to the external field $\phi$. For our concrete calculation we work (as above) with a point-like hotspot relative to which the qubit is at rest and is displaced by $\bfx_\ssQ$. 

We ask in particular how the qubit responds to its proximity to the thermal hotspot, given that its interactions with the hotspot are filtered through the intermediary field $\phi$. Our focus is on the late-time thermalization behaviour; a time-scale that varies inversely with the qubit-field coupling, and so lies beyond the reach of naive perturbation theory. Following \cite{Kaplanek:2019dqu, Kaplanek:2019vzj, Kaplanek:2020iay} we use Open-EFT techniques to access this late-time limit, with the goal of providing a point of comparison for similar calculations in more complicated black-hole and cosmological geometries. We also explore in this simple setting how qubit thermalization depends on an interplay between the strength of its couplings and distance from the hotspot.  For simplicity, throughout this section we take the hotspot-localized $\phi$ self-interaction coupling to vanish: $\lambda=0$.

\subsection{Qubit evolution equations}

The free Hamiltonian for the qubit-field system is assumed to be described by the Hamiltonian
\be
 {H}_{\mathrm{tot}}(t) = {H}_{\ssS}(t) \otimes \boldsymbol{I}    +  \cI  \otimes  \boldsymbol{H}_{0}    +  {H}_{\mathrm{int}}^{\ssQ}(t) \,,
\ee
where ${H}_{\ssS}(t)$ is the Schr\"odinger-picture Hamiltonian for the fields $\phi$ and $\chi^a$ described in \S\ref{sec:setup}, $\cI$ and $\boldsymbol{I}$ are (respectively) the unit operators acting on the Hilbert space for these fields and on the Hilbert space of the qubit. $\boldsymbol{H}_{0}$ denotes the free $2\times 2$ qubit Hamiltonian, and is assumed (in its rest frame) to be
\bea
\boldsymbol{H}_{0}  & = & \frac{\omega}{2} \,\boldsymbol{\sigma_3} \ = \ \frac{\omega}{2} \left[ \begin{matrix} 1 & 0 \\ 0 & -1 \end{matrix} \right] \ ,
\eea
where $\omega$ denotes the splitting between its two levels. 

$H^\ssQ_{\rm int}$ describes the Schr\"odinger-picture qubit-field interaction, in which the qubit couples only to the field $\phi$ evaluated at the local qubit position, taken to be at rest relative to the hotspot and displaced from it by $\bfx_\ssQ$. This interaction is chosen to drive transitions between the qubit levels,
\be
 {H}_{\mathrm{int}}^{\ssQ}(t)=   \hat\lambda_{\ssQ}(t) \;  {\phi}_{\ssS}(\bfx_\ssQ) \otimes \cI_- \otimes \boldsymbol{\sigma_1} \quad
\hbox{ where} \; \boldsymbol{\sigma_1} = \left[  \begin{matrix} 0 & 1 \\ 1 & 0 \end{matrix} \right] \,,
\ee
where $\cI_-$ is the unit matrix in the $\chi^a$ sector of the Hilbert space, and the dimensionless coupling parameter $\hat\lambda_\ssQ$ is assumed to be small so as to justify treating the qubit-field interaction perturbatively. 

We imagine the qubit-field coupling to be turned on suddenly at time $t = t_0$,
\be \label{qubitTurnon}
  \hat\lambda_\ssQ(t) = \lambda_{\ssQ} \, \Theta(t - t_0) 
\ee
with $t_0 > 0$ so that switch-on occurs after the fields have already begun to interact. Our focus is not on the transients associated with this turn-on, and instead on the qubit's late-time approach to equilibrium and on how this approach depends on the other scales of the problem such as the distance $|\bfx_\ssQ|$ between the qubit and the hotspot. 

To compute the qubit evolution we adopt the interaction picture, with the interaction Hamiltonian including only $H^\ssQ_{\rm int}(t)$. In this picture the interactions between the $\phi$ and $\chi^a$ fields are all regarded as being within the `unperturbed' Hamiltonian. Interaction picture evolution of the fields $\phi$ and $\chi^a$ is therefore the same as what was considered Heisenberg-picture evolution in the absence of the qubit, and so is given by the same equations --- {\it i.e.}~eqs.~\pref{heis1} and \pref{heis2} --- that were solved in \cite{Hotspot}.

In this interaction picture the system state evolves purely due to the qubit-field interaction and so does not evolve at all until the time $t = t_0$ when these turn on. As a consequence the system's density matrix at $t = t_0$ remains unchanged from its initially uncorrelated configuration at $t = 0$:
\be
R_{\ssS}(t_0) = R_{\ssS}(0) = \rho_0 \otimes \varrho(0) \,,
\ee
where $\varrho(0)$ denotes the initial qubit state and $\rho_0$ is the field state given in \pref{initialstate}, in which the $\phi$ sector starts in its vacuum while the $\chi^a$ are prepared in a thermal state.

The evolution of the system's state for $t > t_0$ is given in the interaction picture by
\be \label{RevoInt}
\frac{\partial R(t)}{\partial t} = - i \Bigl[ {V}_{\mathrm{tot}}(t), R(t) \Bigr] \ ,
\ee
where $V_{\rm tot}(t)$ is the interaction-picture interaction Hamiltonian for the field-qubit system coupling is defined by
\be  \label{QubitVint}
{V}_{\mathrm{tot}}(t) \ = \ \bigg( {U}^{\ast}(t,0) \otimes e^{+ i \boldsymbol{H}_{0} t} \bigg) \; {H}_{\mathrm{int}}(t) \; \bigg( {U}(t,0) \otimes e^{- i \boldsymbol{H}_{0} t} \bigg) \ = \ \hat\lambda_{\ssQ}(t) \; {\phi}_{\ssI}(t,\mathbf{x}_{\ssQ}) \otimes  \boldsymbol{\mfm}(t)
\ee
where 
\be
  {U}(t,s) = \mathcal{T}\,\exp\left( - i \int_s^t \exd \tau\; {H}_{\ssS}(\tau) \right) \,,
\ee
is the evolution operator for the fields $\phi$ and $\chi^a$. The second equality in eq.~\pref{QubitVint} uses the time-evolution property for interaction-picture fields: $\phi_\ssI(t,\bfx) =  {U}^{\ast}(t,0) {\phi}_{\ssS}(\mathbf{x}_{\ssQ}) {U}(t,0)$, where $\phi_\ssS$ is the Schr\"odinger-picture operator. Notice that the interaction-picture field $\phi_\ssI$ appearing in \pref{QubitVint} is precisely the same as what was called the Heisenberg-picture field $\phi_\ssH$ in \S\ref{sec:setup}, since that section did not include the field-qubit interaction being considered here. As a result the Wightman function for $\phi_\ssI(t,\bfx)$ is given by the expressions \pref{fullW}, \pref{curlySanswer} and \pref{curlyEanswer} computed in \cite{Hotspot}.  Finally, the matrix $\boldsymbol{\mfm}(t)$ is similarly defined as
\be
\boldsymbol{\mfm}(t)   :=   e^{+ i \boldsymbol{H}_{0} t} \boldsymbol{\sigma_1} e^{- i \boldsymbol{H}_{0} t} \ = \ \left[ \begin{matrix} 0 & e^{+ i \omega t} \\ e^{- i \omega t} & 0 \end{matrix} \right] \ .
\ee

\subsection{Tracing out the fields and late-time qubit evolution}

Since our goal is to follow only the dynamics of the qubit keeping track of the entire density matrix $R(t)$ carries too much information. For qubit measurements it suffices instead to follow the evolution of the qubit's reduced interaction-picture density matrix 
\be
\varrho(t)   :=   \TrAB\left[ R(t) \right] \ ,
\ee
since this carries the information required to predict measurements in this sector. At first sight the evolution of this reduced density matrix is obtained simply by tracing over \pref{RevoInt}, though the resulting equation has the disadvantage that its right-hand side is not expressed purely in terms of $\varrho$ without reference to the field part of the system's density matrix. 

A more useful equation would compute the evolution of both $\varrho(t)$ and the density matrix for the field sector, and then use the evolution equations to eliminate the field density matrix as a function of $\varrho$. Doing so is a solved problem in the theory of open quantum systems, and the resulting self-contained equation for the evolution of $\varrho(t)$  is the so-called Nakajima-Zwanzig equation \cite{Nak, Zwan} (for a review see {\it e.g.}~\cite{EFTBook}). The logic of its derivation is to exploit the linearity of the evolution equation \pref{RevoInt} and to compute how it commutes with a projection super-operator $\mathscr{P} \{ \cdot \}$ that maps operators acting on the full Hilbert space onto operators acting only within the qubit (in this case) sector, defined in such a way as to ensure that $\mathscr{P}[R(t)] = \varrho(t)$. 

The idea then is to compute what \pref{RevoInt} predicts for both  $\mathscr{P}\{ R(t) \}$ and for its complement $\mathscr{Q}\{ R(t) \}$, where  $\mathscr{Q}\{ R(t) \} = \cI - \mathscr{P}\{ R(t) \}$. The result is a coupled set of linear evolution equations, which can be formally integrated to obtain $\mathscr{Q}\{ R \}$ as a function of time and of $\varrho(t)$. Substituting the result back into the equation for $\mathscr{P}\{R(t)\}$ provides the equation we seek. Because of the elimination of $\mathscr{Q}\{R(t)\}$ the equation for $\varrho(t)$ that results is generically nonlocal in time, making $\partial_t\varrho(t)$ depend on $\varrho(t)$ but also on the entire history $\varrho(s)$ for $s < t$. 

The application of this equation to qubit systems in various environments is studied in detail in \cite{Kaplanek:2019dqu, Kaplanek:2019vzj, Kaplanek:2020iay} so we here simply quote what the Nakajima-Zwanzig equation gives for $\varrho(t)$ in the present instance. The result can be computed explicitly as a series in the coupling $\lambda_\ssQ$, and when evaluated for $t > t_0$ to second order in $\lambda_{\ssQ}$ takes the form
\bea\label{qubitNZ} 
\partial_t \boldsymbol{\varrho}(t) & = &  - i   \bigg[ \frac{\lambda_{\ssQ}^2\, \omega_{\rm ct}}{2}\, \boldsymbol{\sigma_3} , \boldsymbol{\varrho}(t)  \bigg]  \\
& \ & \qquad  +  \lambda_{\ssQ}^2 \int_{t_0}^t \exd s\; \bigg( \Bigl[ \boldsymbol{\mfm}(s) \boldsymbol{\varrho}(s) ,  \boldsymbol{\mfm}(t)  \Bigr] \cW(t , s) + \Bigl[ \boldsymbol{\mfm}(t) , \boldsymbol{\varrho}(s) \boldsymbol{\mfm}(s)  \Bigr] \cW^{\ast}(t, s)  \bigg) \notag
\eea
where $\omega_{\rm ct}$ of the first term is a counter-term for the qubit frequency, which is written $\omega = \omega_{\rm phys} + \lambda_\ssQ^2\omega_{\rm ct}$ with $\omega_{\rm ct}$ chosen to ensure that $\omega_{\rm phys}$ remains the physically measured qubit energy to the order we work in $\lambda_\ssQ$. This shift is required because corrections to the qubit energy arise at order $\lambda_\ssQ^2$, which $\omega_{\rm ct}$ is chosen to cancel.\footnote{As it happens these corrections also diverge and so $\omega_{\rm ct}$ provides the counterterm that cancels this divergence.} Because $\omega_{\rm ct}$ is of order $\lambda_\ssQ^2$, within the interaction picture it is included into the perturbing Hamiltonian by writing
\be
{V}_{\mathrm{tot}}(t)  =  \hat\lambda_{\ssQ}(t) \; {\phi}_{\ssI}(t,\mathbf{x}_{\ssQ})   \otimes \boldsymbol{\mfm}(t)   +      \frac{\hat\lambda_{\ssQ}^2(t)  \, \omega_{\rm ct}}{2} \;\cI  \otimes \boldsymbol{\sigma_{3}} \ . 
\ee

The convolution in the second line of \pref{qubitNZ} reveals how $\partial_t \varrho$ depends on the previous history of the qubit's evolution, and the kernel $\cW(t_1, t_2)$ appearing in this convolution is the Wightman function for the field $\phi_\ssI$ --- {\it i.e.}~precisely the quantity quoted above in \S\ref{sec:setup} that is calculated explicitly in \cite{Hotspot} --- evaluated at two points along the qubit world-line:
\be  \label{heiscorr3}
\cW(t_1, t_2)  :=  \mathrm{Tr}\Bigl[ \phi_{\ssI}(t_1, \bfx_\ssQ) \phi_{\ssI}(t_2, \bfx_\ssQ)  \rho_0 \Bigr] \,.
\ee

Because \pref{qubitNZ} is a $2 \times 2$ matrix equation it looks harder to solve than it really is. In particular, the properties $\mathrm{tr} \,\boldsymbol{\varrho}  = 1$ and $\boldsymbol{\varrho}^{\dagger}  = \boldsymbol{\varrho}$ can be used to eliminate $\varrho_{22} = 1 - \varrho_{11}$ and $\varrho_{21} = \varrho_{12}^{\ast}$ from these equations, so it suffices to know how $\varrho_{12}$ and $\varrho_{11}$ evolve. Using \pref{qubitNZ} to evaluate the evolution for these two components reveals that they decouple from one another, and so evolve independently with 
\bea\label{11eqNZ}
\frac{\pd \varrho_{11}(t)}{\pd t} & = & 2 \lambda_{\ssQ}^2 \int_{t_0}^{t} \exd s\;   \bigg( \mathrm{Re}[\cW(t, s)] \cos(\omega [t - s]) + \mathrm{Im}[\cW(t, s)] \sin(\omega [t - s]) \bigg) \notag  \\
&& \quad \quad \quad \quad \quad - 4  \lambda_{\ssQ}^2  \int_{t_0}^{t} \exd s\;   \mathrm{Re}[\cW(t, s)] \cos(\omega [t - s]) \varrho_{11}(s) 
\eea
and (for $t > t_0$)
\bea  \label{12eqNZ} 
\frac{\pd \varrho_{12}(t)}{\pd t} & = & - i   \lambda_{\ssQ}^2\, \omega_{\rm ct} \;\varrho_{12}(t) - 2 \lambda_{\ssQ}^2  \int_{t_0}^{t} \exd s\;   \mathrm{Re}[\cW(t, s)] e^{+ i \omega [t - s]} \varrho_{12}(s) \notag \\
& \ & \quad \quad \quad \quad \quad + 2  \lambda_{\ssQ}^2  e^{+ 2 i \omega t } \int_{t_0}^{t} \exd s\;  \mathrm{Re}[\cW(t, s)] e^{- i \omega [t - s]} \varrho^{\ast}_{12}(s) \,.
\eea

The evolution equations normally used (for instance in \cite{Sciama:1981hr}) when perturbatively treating Unruh-DeWitt detectors are obtained from \pref{11eqNZ} and \pref{12eqNZ} by replacing all appearances of $\varrho_{ij}(t)$ on the right-hand side with the initial condition $\varrho_{ij}(t_0)$. (In particular these terms would all vanish if the intial state was the ground state, which corresponds to $\varrho_{22}(t_0) = 1$ with all others zero in the present notation.) Indeed replacing $\varrho_{ij}(t)$ with $\varrho_{ij}(t_0)$ seems very reasonable at first sight because the difference between $\varrho_{ij}(t)$ and $\varrho(t_0)_{ij}$ is higher order in $\lambda_\ssQ$ and so straight-up perturbation theory should drop this difference. Eqs.~\pref{11eqNZ} and \pref{12eqNZ} are nonetheless better approximations at late times and disagree with naive perturbation theory precisely because perturbative methods break down at late times.

Our intended application of these expressions is to understand whether (and how quickly) the qubit thermalizes due to its indirect interaction with the hotspot, and we have no interest in the transients associated with the turn-on of couplings. This makes very late times our focus of interest, and so we restrict our attention to qubit positions and times that satisfy
\be
t \ > \ t_0 \ > \ |\bfx_{\ssQ}| \,,
\ee
where $t_0$ is the turn-on time for the qubit interaction appearing in \pref{qubitTurnon}. The choice $t_0 > |\bfx_{\ssQ}|$ ensures that the outgoing wave caused by the $t = 0$ turn-on of the $\phi$-$\chi^a$ field couplings have had time to have passed the location of the qubit. In practice we choose times in the far future for which the full correlation function \pref{fullW} (without the step- and delta-functions) can be used. Since in this limit $\cW(t,s) = \cW(t-s)$ it is convenient to define $\widetilde\cW$ by
\be  \label{Wtildedef}
\widetilde{\cW}(\tau)  : = \cW(\tau,0) = \mathscr{S}(\tau, \bfx_{\ssQ} ; 0, \bfx_{\ssQ} ) |_{\lambda = 0} + \mathscr{E}_\beta(\tau, \bfx_{\ssQ} ; 0, \bfx_{\ssQ} ) |_{\lambda = 0} \,,
\ee
where $\mathscr{S}$ and $\mathscr{E}_\beta$ are the functions defined in \pref{curlySanswer} and \pref{curlyEanswer}. Explicitly
\bea \label{Wttinv}
\widetilde{\cW}(\tau)  
& = &  -  \frac{1}{4 \pi^2 (\tau - i \delta)^2}  + \frac{ 2 \epsilon^2}{\tilde{g}^2 |\bfx_{\ssQ}|^2} \bigg[ I_{-}(\tau + 2 |\bfx_{\ssQ}|, c_0) - I_{-}(\tau, c_0) - I_{+}( \tau , c_0 ) + I_{+}( \tau-  2|\bfx_{\ssQ}| , c_0) \bigg] \nn \\
& \ & \qquad + \frac{\epsilon}{8 \pi^2 |\bfx_{\ssQ}|^2} \bigg[ - \frac{1}{\tau + 2 |\bfx_{\ssQ}| - i \delta} + \frac{1}{\tau -2 |\bfx_{\ssQ}| - i \delta } \bigg]  \\
& \ & \qquad \qquad -  \frac{32 \pi^2 \epsilon^4}{ \tilde{g}^4 |\bfx_{\ssQ}|^2} \bigg[ I_{-}( \tau , c_0 ) + I_{+}( \tau , c_0 )  \bigg] - \frac{\epsilon^2}{4 \pi^2 |\bfx_{\ssQ}|^2 ( \tau - i \delta )^2} \nn \\
& \ &  \qquad \qquad \qquad +  \frac{2\epsilon^2}{\tilde{g}^2 |\bfx_{\ssQ}|^2} \bigg[ \Phi\bigg( e^{ - \tfrac{2\pi (\tau - i \delta)}{\beta}}, 1 , \frac{c_0\beta}{2\pi} \bigg) + \Phi\bigg( e^{ + \tfrac{2\pi (\tau - i \delta)}{\beta}}, 1 , \frac{c_0\beta}{2\pi} \bigg)  - \frac{2\pi}{c_0\beta} \bigg] \nn
\eea
with the functions $\Phi$ and $I_\pm(\tau,c_0)$ as defined in \pref{IpmDef} and \pref{PhiDef}. The parameter $c_0$ here denotes 
\be
c_0 \ := \ c\; |_{\lambda = 0} \ = \ \frac{16\pi^2 \epsilon}{\tilde{g}^2}
\ee
which arises because we set $\lambda =0$ in this section ({\it c.f.}~\pref{cdef}).

After a change of variable $s \to t - s$ on the right-hand side of (\ref{11eqNZ}) and (\ref{12eqNZ}), and using the symmetry $\widetilde{\cW}^{\ast}(\tau) = \widetilde{\cW}(-\tau)$ of the Wightman functions that follows from the hermiticity of the field ${\phi}$, we get the evolution equations in the form we ultimately solve
\bea  \label{11eqNZ3}
\frac{\pd \varrho_{11}(t)}{\pd t} & = & \lambda_{\ssQ}^2 \int_{-(t-t_0)}^{t-t_0} \exd s\; \widetilde{\cW}(s) \, e^{- i \omega s} - 4 \lambda_{\ssQ}^2 \int_{0}^{t-t_0} \exd s\;  \mathrm{Re}[\widetilde{\cW}(s)] \cos(\omega s) \varrho_{11}(t-s)
\eea
and
\bea \label{12eqNZ3}
\frac{\pd \varrho_{12}(t)}{\pd t} & = & - i \lambda_{\ssQ}^2\, \omega_{\rm ct} \varrho_{12}(t) - 2 \lambda_{\ssQ}^2 \int_{0}^{t-t_0} \exd s\; \mathrm{Re}[\widetilde{\cW}(s)] \, e^{+ i \omega s} \varrho_{12}(t-s)  \\
& \ & \qquad \qquad \qquad \quad \quad + 2  \lambda_{\ssQ}^2 \, e^{+ 2 i \omega t } \int_{0}^{t-t_0} \exd s\; \mathrm{Re}[\widetilde{\cW}(s)]\,  e^{- i \omega s} \varrho^{\ast}_{12}(t-s) \,.\notag
\eea

\subsection{The Markovian limit}

Equations \pref{11eqNZ3} and \pref{12eqNZ3} are in general difficult to solve, largely due to the convolutions appearing on their right-hand sides. We seek here approximate solutions in the special situation where the kernel $\widetilde\cW(s)$ varies over some time-scale $\tau_c$, say, that is much shorter than the scale $\tau_\rho$ over which $\varrho(t-s)$ varies. In such a case the integrands of eqs.~\pref{11eqNZ3} and \pref{12eqNZ3} can be usefully expanded in powers of $s$, 
\be \label{taylorMarkov}
\boldsymbol{\varrho}(t - s) \simeq \boldsymbol{\varrho}(t) - s\, \partial_t{\boldsymbol{\varrho}}(t) + \ldots \ , 
\ee
with successive terms suppressed by powers of $\tau_c/\tau_\rho$ after the integration over $s$ is performed. Once derivatives of $\varrho$ can be neglected then equations \pref{11eqNZ3} and \pref{12eqNZ3} become Markovian because they give $\partial_t \varrho(t)$ directly in terms of $\varrho(t)$ (without a convolution over earlier times) and can be integrated with little difficulty.  

Closer inspection of (\ref{Wttinv}) indeed reveals its last terms, involving the function $\Phi(z,s,a)$, to be peaked --- with exponential fall-off (see Appendix \ref{App:curlyEpert}) --- about $s =0$, with a width $\tau_c \sim \beta$, suggesting that a Markovian limit might apply for evolution over times much larger than $\beta$. The other terms -- coming from $\mathscr{S}$ in \pref{curlySanswer} -- are trickier because they fall off more slowly (like a power-law rather than exponentially). Because this fall-off is slower, care is required to justify the Markovian for these terms.

Our strategy for solving for qubit evolution is to assume that a Markovian regime exists, use it to identify whether the qubit thermalizes, and then justify {\it ex post facto} that the Markovian approximation is justified for the parameter range that gives thermalization. This suffices for our purposes of establishing that thermalization occurs, but does not exclude the Markovian regime having a broader domain of validity than we identify here. 

Keeping only the leading term of the expansion \pref{taylorMarkov} in \pref{11eqNZ3} and \pref{12eqNZ3} gives the following approximate evolution equations,
\be
\frac{\pd \varrho_{11}(t)}{\pd t}   \simeq   \lambda_{\ssQ}^2 \int_{-(t-t_0)}^{t-t_0} \exd s\; \widetilde{\cW}(s) \, e^{- i \omega s} - 4 \lambda_{\ssQ}^2\, \varrho_{11}(t) \int_{0}^{t-t_0} \exd s\;  \mathrm{Re}[\widetilde{\cW}(s)] \cos(\omega s) 
\ee
and
\bea
\frac{\pd \varrho_{12}(t)}{\pd t} & \simeq & - i \lambda_{\ssQ}^2 \omega_{\rm ct} \varrho_{12}(t) - 2 \lambda_{\ssQ}^2\,  \varrho_{12}(t) \int_{0}^{t-t_0} \exd s\; \mathrm{Re}[\widetilde{\cW}(s)] \, e^{+ i \omega s}  \\
&& \qquad \qquad \qquad \quad \quad + 2  \lambda_{\ssQ}^2 \,e^{+ 2 i \omega t } \varrho^{\ast}_{12}(t)  \int_{0}^{t-t_0} \exd s\; \mathrm{Re}[\widetilde{\cW}(s)] \, e^{- i \omega s}\,, \notag
\eea
and these can be simplified even further if we focus on times $t - t_0 \gg \beta\,, \omega^{-1}$, since we can then with small error replace $t-t_0 \to \infty$ in the limits of integration. For the thermal part of the Wightman function the error made with this replacement is exponentially small due to the exponential falloff of the $\mathscr{E}_\beta$ term in $\widetilde{\cW}$. The slower fall-off of $\mathscr{S}$ implies the error in doing so can instead in principle involve inverse powers of $\omega(t-t_0)$.  

Under the above circumstances the evolution equations take the form we shall integrate
\be\label{11eqNZ4}
\frac{\pd \varrho_{11}(t)}{\pd t}   \simeq   \lambda_{\ssQ}^2 \,\cR - 2 \lambda_{\ssQ}^2 \, \cC \, \varrho_{11}(t) 
\ee
and
\be \label{12eqNZ4} 
\frac{\pd \varrho_{12}(t)}{\pd t}   \simeq   -  \lambda_{\ssQ}^2 \Bigl[ \cC + i ( \omega_{\rm ct} +  \cD ) \Bigr] \varrho_{12}(t)   +    \lambda_{\ssQ}^2 \, e^{+ 2 i \omega t } \, \cC\, \varrho^{\ast}_{12}(t)  \,.
\ee
Here the coefficients $\cC$, $\cD$ and $\cR$ are defined by
\be \label{Cqubitdef}
\cC = 2 \int_{0}^{\infty} \exd s\;  \mathrm{Re}[\widetilde{\cW}(s)] \cos(\omega s)  \,, \quad
\cD = 2 \int_{0}^{\infty} \exd s\;  \mathrm{Re}[\widetilde{\cW}(s)] \sin(\omega s)  
\ee
and
\be \label{Rqubitdef} 
\cR   =   \int_{-\infty}^{\infty} \exd s\; \widetilde{\cW}(s)\, e^{- i \omega s} \,.
\ee

These integrals are evaluated explicitly in Appendix \ref{App:UDWintegrals} using the form of $\widetilde\cW$ given in eq.~\pref{Wttinv}. For instance, the result for $\cR$ when $\omega > 0$ is given by
\be \label{cRIntegralResult}
\cR = \frac{\tilde{g}^2 \omega }{32 \pi^3 |\bfx_{\ssQ}|^2 \left[ \big( \frac{\tilde{g}^2 \omega}{16 \pi^2 \epsilon} \big)^2 + 1 \right] \left( e^{\beta \omega} - 1 \right)} \,.
\ee
Notice that this expression vanishes in the zero-temperature limit, and also vanishes as $\tilde g^2 \to 0$ despite the presence of terms independent of $\tilde g^2$ in expression \pref{curlySanswer} for $\mathscr{S}$. As discussed some time ago \cite{Sciama:1981hr} this is a consequence of having $\omega > 0$ because it relies on the vacuum spectral density having no support for positive frequencies. It is what prevents the vacuum from spontaneously exciting a static qubit initially prepared in its ground state.

The expressions for $\cC$ and $\cD$ are equally explicit, though slightly more complicated:
\be \label{cCIntegralResult}
\cC = \frac{\omega}{4 \pi} \left[ 1 - \frac{\tilde{g}^2}{16 \pi^2 |\bfx_\ssQ|^2 } \left(\frac{ 1 - \cos\big( 2 \omega |\bfx_{\ssQ}| \big) - \frac{\tilde{g}^2\omega^2}{16 \pi^2}  + \frac{\tilde{g}^2 \omega}{16 \pi^2 \epsilon}  \sin\big( 2 \omega |\bfx_{\ssQ}|  \big) - \mathrm{coth} \big( \frac{\beta \omega}{2} \big)}{ \big( \frac{\tilde{g}^2 \omega}{16 \pi^2 \epsilon} \big)^2  +1} \right)   \right] \,,
\ee
and
\bea \label{Danswer}
\cD & = & \frac{\omega}{2\pi^2} \bigg[ 1 + \frac{\epsilon^2}{|\bfx_{\ssQ}|^2} \bigg]  \log\left( \frac{\omega}{\Lambda} \right) - \frac{\omega \epsilon^2}{2\pi^2 |\bfx_{\ssQ}|^2} \cdot \frac{1}{ (\omega/c_0)^2 + 1 } \log\left( \frac{\omega}{c_0} \right) \nn \\
& \ & \quad -  \frac{\tilde{g}^2\omega}{32 \pi^4 |\bfx_{\ssQ}|^2}  \cdot \frac{(\omega/c_0)}{(\omega/c_0)^2+ 1} \cdot \bigg[ \mathrm{Ci}\big(2 |\bfx_{\ssQ}| \omega\big) \sin( 2 |\bfx_{\ssQ}| \omega ) - \mathrm{Si}\big(2 |\bfx_{\ssQ}| \omega\big) \cos( 2 |\bfx_{\ssQ}| \omega ) \bigg]  \\
& \ & \quad  + \frac{\tilde{g}^2 \omega}{32 \pi^4 |\bfx_{\ssQ}|^2} \cdot \frac{1}{(\omega/c_0)^2 + 1} \cdot \bigg\{ \mathrm{Ci}\big(2 |\bfx_{\ssQ}| \omega\big) \cos( 2 |\bfx_{\ssQ}| \omega ) + \mathrm{Si}\big(2 |\bfx_{\ssQ}| \omega\big) \sin( 2 |\bfx_{\ssQ}| \omega )  \nn \\
& \ & \qquad \qquad \qquad \qquad - e^{ - 2 |\bfx_{\ssQ}| c_0} \mathrm{Ei}\big(2|\bfx_{\ssQ}|c_0\big)  - \log\left( \frac{\omega}{c_0} \right) - \psi^{(0)}\left( \frac{\beta c_0}{2\pi} \right) + \mathrm{Re}\left[ \psi^{(0)}\left( i \frac{\beta \omega}{2\pi} \right) \right] - \frac{\pi}{\beta c_0} \bigg\} \nn \ .
\eea
where 
\be
   \mathrm{Si}(z): = \int_0^z \exd t\; \frac{\sin t}{t} \,, \quad 
   \mathrm{Ci}(z) := - \int_{z}^{\infty} \exd t\; \frac{\cos t}{t} \quad \hbox{and} \quad
   \mathrm{Ei}(z) := - \int_{-z}^{\infty} \exd t\; \frac{e^{-t}}{t}
\ee 
are standard functions and $\psi^{(0)}(z) := \Gamma'(z) / \Gamma(z)$ is the digamma function. 

Before exploring the late-time qubit evolution we pause to remark on several noteworthy features of these expressions.
\begin{itemize}
\item
As mentioned earlier, for qubits started in their ground state the initial conditions are $\varrho_{11}(t_0) = \varrho_{12}(t_0) = 0$ and so if $\varrho_{11}(t)$ were replaced by its initial condition $\varrho_{11}(t_0)$ on the right-hand side of \pref{11eqNZ4} the evolution equation would reduce to the result obtained by straightforward application of perturbation theory:
\be\label{11eqNZ4a}
\frac{\pd \varrho_{11}(t)}{\pd t}   \simeq   \lambda_{\ssQ}^2 \,\cR   \qquad (\hbox{perturbative evolution from ground state}) \,.
\ee
This agrees with early calculations for Unruh-DeWitt detectors \cite{Sciama:1981hr}, which identified $\cR$ as the qubit's excitation rate. As we see below, this rate differs from the thermalization rate calculated at late times (these rates also differ for Unruh-DeWitt detectors in nontrivial spacetimes \cite{Kaplanek:2019dqu, Kaplanek:2019vzj, Kaplanek:2020iay}). 
\item 
The parameter $\Lambda$ appearing in \pref{Danswer} is an ultraviolet regulator whose presence shows that the function $\cD$ diverges. It does so because of the singular behaviour of $\cW_\beta(\tau)$ as $\tau \to 0$. As is usual for UV divergences, this is renormalized into the value of a parameter, in this case the counter-term $\omega_{\rm ct}$, as can be seen from the fact that $\omega_{\rm ct}$ and $\cD$ only appear in eq.~\pref{12eqNZ4} and only do so there together as the sum $\omega_{\rm ct} + \cD$, and so $\omega_{\rm ct}$ can be chosen to cancel any $\bfx_\ssQ$-independent part of $\cD$. 

At face value the requirement of $\bfx_\ssQ$-independence might appear to be a problem because \pref{Danswer} contains a $\bfx_\ssQ$-dependent divergence. However this 
$\bfx_\ssQ$-dependence drops out in the regime where the qubit is macroscopically far from the hotspot, $\epsilon \ll \bfx_\ssQ$, and in this regime $\bfx_\ssQ$-independence of the divergence is guaranteed by the Hadamard property of the Wightman function in the coincidence limit (see Appendix \ref{App:Hadamard}).

\end{itemize}

\subsection{Equilibrium and its approach}

Eqs.~\pref{11eqNZ4} and \pref{12eqNZ4} are relatively straightforward to integrate, and for simplicity we choose parameters to be in the `non-degenerate' regime, for which the initial qubit splitting $\omega$ is much larger than any of the $\cO(\lambda_\ssQ^2)$ corrections to this splitting; or, in practice:
\be\label{nondegen}
\left| \frac{\lambda_{\ssQ}^2 \cC}{\omega} \right| \ll 1 \qquad \mathrm{and} \qquad \left| \frac{\lambda_{\ssQ}^2 \cD}{\omega} \right| \ll 1 \,.
\ee

\subsubsection{Solutions}

The explicit solutions to \pref{11eqNZ4} and \pref{12eqNZ4} in this regime are
\be  \label{11sol1}
\varrho_{11}(t) = \frac{\cR}{2\cC} + \bigg[  \varrho_{11}(t_0) - \frac{\cR}{2 \cC} \bigg] e^{ - 2 \lambda_{\ssQ}^2 \cC (t-t_0)}
\ee
and
\be \label{12sol1}
\varrho_{12}(t) \ \simeq \ \bigg[ \varrho_{12}(t_0) + i \varrho_{12}^{\ast}(t_0) \frac{\lambda_{\ssQ}^2 \cC}{2\omega} (e^{2 i \omega t_0} - e^{ 2 i \omega t}) \bigg] e^{ - \lambda_{\ssQ}^2 \cC (t-t_0)} \ \simeq \  \varrho_{12}(t_0) e^{ - \lambda_{\ssQ}^2 \cC (t-t_0)}  \ , 
\ee
where the approximate equality uses \pref{nondegen} to neglect the rapidly oscillating term.\footnote{Equivalently, \pref{nondegen} justifies neglecting the second term on the right-hand side of \pref{12eqNZ4} relative to the first term.} These solutions describe an exponential relaxation towards a static late-time configuration
\be
  \lim_{t \to \infty} \boldsymbol{\varrho}(t) = \boldsymbol{\varrho}_{\star} =  \left[ \begin{matrix} \frac{\cR}{2\cC} & 0 \\ 0 & 1 - \frac{\cR}{2\cC} \end{matrix} \right] \,,
\ee
with a relaxation time that differs for the diagonal and the off-diagonal elements, 
\be \label{decaytimes}
   \tau_{\rm diag} = \frac{1}{2\lambda_\ssQ^2 \cC} \quad \hbox{and} \quad
   \tau_{\rm off-diag} = \frac{1}{\lambda_\ssQ^2 \cC} \,.
\ee

The nonzero diagonal elements of the late-time state $\boldsymbol{\varrho}_\star$ evaluate to $\varrho_{*11} = \cR/(2\cC)$ where eqs.~\pref{cRIntegralResult} and \pref{cCIntegralResult} imply
\be \label{Rover2C}
\frac{\cR}{2 \cC} = \frac{ 1 }{ \left( e^{\beta \omega} - 1 \right) \left\{ \frac{16 \pi^2 |\bfx_{\ssQ}|^2}{\tilde{g}^2} \left[  \big( \frac{\tilde{g}^2 \omega}{16 \pi^2 \epsilon} \big)^2 + 1 \right] + \frac{\tilde{g}^2\omega^2}{16 \pi^2} + \cos\big( 2   \omega |\bfx_{\ssQ}| \big) - 1 - \frac{\tilde{g}^2 \omega}{16 \pi^2 \epsilon}  \sin\big( 2 |\bfx_{\ssQ}| \omega  \big) + \mathrm{coth}\big( \frac{\beta \omega}{2} \big) \right\} } \ .
\ee
This describes a thermal distribution\footnote{This is thermal inasmuch as it is diagonal, normalized and the ratio of probabilities for the two qubit states is given by the Boltzmann relation $\varrho_{11}/\varrho_{22} = e^{-\beta \omega}$.}
\be \label{Rover2CThermal}
\frac{\cR}{2 \cC} \simeq   \frac{ 1 }{ \left( e^{\beta \omega} - 1 \right)   \mathrm{coth}\big( \frac{\beta \omega}{2} \big)  }  = \frac{e^{-\beta\omega/2}}{e^{\beta\omega/2} + e^{-\beta\omega/2}}   = \frac{1}{e^{\beta\omega} + 1}   \,,
\ee
at the hotspot temperature $T = 1/\beta$ provided all of the terms save for the last one can be neglected in the curly braces of the denominator of \pref{Rover2C}. A parameter regime that is sufficient to ensure this (and so also to ensure late-time thermality) therefore jointly asks 
\be \label{parameters}
 \frac{16 \pi^2 |\bfx_{\ssQ}|^2}{\tilde{g}^2} \ll 1 \,, \quad
 \frac{\tilde g^2 \omega^2 |\bfx_{\ssQ}|^2}{16\pi^2 \epsilon^2} \ll 1 \,, \quad \frac{\tilde g^2 \omega^2  }{16\pi^2 } \ll 1 \,, \quad  
 \omega |\bfx_\ssQ| \ll 1  \,, \quad
   \frac{\tilde g^2 \omega^2 |\bfx_{\ssQ}|}{16\pi^2 \epsilon} \ll 1\,.
\ee
These are equivalent to the two independent assumptions
\be \label{parameters2}
 |\bfx_{\ssQ}| \ll \frac{\tilde{g} }{4 \pi} \qquad \hbox{and} \qquad \frac{\tilde g \omega}{4 \pi} \ll \frac{\epsilon}{|\bfx_\ssQ|} \ll 1 \,.
\ee
where the last inequality follows because a qubit being well-separated from the hotspot means that it satisfies $|\bfx_\ssQ | \gg \epsilon$. The first of the conditions in \pref{parameters2} is inconsistent with the requirement $|\bfx_\ssQ | \gg \epsilon$ unless the hotspot coupling also satisfies 
\be \label{thermnonpert}
   \frac{\tilde g}{4\pi \epsilon} \gg 1 \,.
\ee
Notice that these conditions do not yet impose a hierarchy on the size of $\omega/c_0 = \tilde g^2\omega/(16\pi^2 \epsilon)$, since \pref{parameters2} only implies this must satisfy $\omega/c_0 \ll \tilde g/(4\pi \epsilon)$ (which is not very informative given \pref{thermnonpert}). 

In the regime defined by \pref{parameters2} expressions \pref{cRIntegralResult}, \pref{cCIntegralResult} and \pref{Danswer} for $\cR$, $\cC$ and $\cD$ become, approximately,
\be \label{cRIntegralResult2}
\cR \simeq \frac{\tilde{g}^2 \omega }{32 \pi^3 |\bfx_{\ssQ}|^2 \big[  ( \omega/c_0 )^2 + 1 \big] \left( e^{\beta \omega} - 1 \right)} \,,
\ee
\be \label{cCIntegralResult2}
\cC \simeq \frac{\omega}{4 \pi} \left[ 1 + \frac{\tilde{g}^2 \mathrm{coth} \big( \frac{\beta \omega}{2} \big)}{16 \pi^2 |\bfx_\ssQ|^2 \big[ ( \omega / c_0 )^2  +1 \big]} \right] \,,
\ee
and
\bea \label{cDIntegralResult2}
\cD & \simeq & \frac{\omega}{2\pi^2} \bigg[ \log\left( \frac{\omega}{\Lambda} \right) - \frac{\epsilon^2}{|\bfx_{\ssQ}|^2} \cdot \frac{1}{ (\omega/c_0)^2 + 1 } \cdot \log\big( \frac{\omega}{c_0} \big)  \\
& \ & \qquad \qquad \qquad \qquad + \frac{\tilde{g}^2}{16 \pi^2 |\bfx_{\ssQ}|^2} \cdot \frac{1}{(\omega/c_0)^2 + 1} \cdot \bigg\{  \mathrm{Re}\left[ \psi^{(0)}\left( i \frac{\beta \omega}{2\pi} \right) \right]  - \psi^{(0)}\left( \frac{\beta c_0}{2\pi} \right)- \frac{\pi}{\beta c_0} \bigg\} \bigg] \nn \ .
\eea
Notice that the divergent part of $\cD$ no longer depends on $|\bfx_{\ssQ}|$ as a consequence of dropping terms suppressed by $\epsilon/|\bfx_\ssQ|$. As mentioned earlier, the divergent part of the correlation function is guaranteed to be $\bfx_\ssQ$-independent for $|\bfx_\ssQ| \gg \epsilon$ as a general consequence of its Hadamard-type singularity structure, as argued in Appendix \ref{App:Hadamard}. 

\subsubsection{Validity of the Markovian approximation}

We close by circling back to check the validity of the Markovian approximation used in transforming eqs.~\pref{11eqNZ3} and \pref{12eqNZ3} into \pref{11eqNZ4} and \pref{12eqNZ4}. This can be done by evaluating the size of the leading subdominant term in the expansion of \pref{taylorMarkov}, and demanding that it be parametrically smaller than the dominant term. 

As shown in detail in \cite{Kaplanek:2019dqu, Kaplanek:2019vzj, Kaplanek:2020iay} the conditions for this to be true can be expressed in terms of the integrals $\cC$ and $\cD$ defined in \pref{Cqubitdef}, with the Markovian approximation being valid when the following four quantities are all small:
\be \label{MarkovianBounds}
\left| \lambda_{\ssQ}^2 \frac{\exd \cC}{\exd \omega} \right| \ll 1 \ \ , \qquad \left| \lambda_{\ssQ}^2 \frac{\exd \cD}{\exd \omega} \right| \ll 1 \ \ , \qquad \left| \frac{\omega}{\cC} \; \frac{\exd \cC}{\exd \omega} \right| \ll 1 \quad \mathrm{and} \qquad \left| \frac{\omega}{\cC} \;  \frac{\exd \cD}{\exd \omega} \right| \ll 1 \ .
\ee
The implications of these four conditions --- and of conditions \pref{nondegen} --- are worked out in detail in Appendix \ref{App:MarkovControl} for different parts of parameter space consistent with  the asymptotic expressions \pref{cCIntegralResult2} and \pref{cDIntegralResult2}. 

The resulting constraints on the parameters are listed in Tables \ref{nonDegenTable1}, \ref{MarkovianTable1} and \ref{MarkovianTable2}, where different rows correspond to different assumptions for the relative sizes of the parameters $\beta \omega$ and $\beta c_0$. Although the first two conditions of \pref{MarkovianBounds} can be satisfied simply by making $\lambda_\ssQ$ sufficiently small, the same is not so for the second two. Taken together these tables show that validity of the Markovian approximation requires the additional three conditions 
\be \label{parametersM}
\frac{\lambda_\ssQ^2}{4\pi} \ll 1 \,, \qquad \omega \ll c_0  = \frac{16\pi^2 \epsilon}{\tilde g^2} \qquad  \mathrm{and} \qquad \beta \omega \ll 1  \,,
\ee
above and beyond those of \pref{parameters2}. 

In particular, it happens that the requirement $\beta\omega \ll 1$ --- that follows from the conditions listed in Table \ref{MarkovianTable2} --- is quite powerful and restrictive. In particular, the high-temperature condition $\beta\omega \ll 1$ ensures that Markovian relaxation leads to a largely $\beta$-independent and maximally mixed qubit distribution, with 
\be
  \varrho_{11} \simeq \varrho_{22}\simeq \frac{1}{2}  \,.
\ee

\section*{Acknowledgements}
We thank Sarah Shandera for making the suggestion that got this project started, and KITP Santa Barbara for hosting the workshop (during a pandemic) that led us to think along these lines. (Consequently this research was supported in part by the National Science Foundation under Grant No. NSF PHY-1748958.) CB's research was partially supported by funds from the Natural Sciences and Engineering Research Council (NSERC) of Canada. Research at Perimeter Institute is supported in part by the Government of Canada through the Department of Innovation, Science and Economic Development Canada and by the Province of Ontario through the Ministry of Colleges and Universities.

\appendix

\section{Asymptotic forms and perturbative limits}
\label{App:modesumfull}

In this appendix we reproduce various limits of the Wightman function given in the main text in eq.~\pref{fullW} with \pref{curlySanswer} and \pref{curlyEanswer}, following the discussion of \cite{Hotspot}. The main limits we explore are the coincident limit relevant to studying its Hadamard properties, and the large-separation limit relevant to the Markovian approximation used in the main text. This last limit also controls the perturbative limit as $\tilde g^2 \to 0$, and so the formulae we derive also confirm expression \pref{pertcorr} of the main text as correctly describing the perturbative limit. 

\subsection{Coincidence limit and Hadamard form}
\label{App:Hadamard}

We first study the Wightman function's coincident limit, doing so by comparing two spacetime points that are coincident in space, $\bfx = \bfx'$ at a distance $r := |\bfx| = |\bfx'|$ from the hotspot, but are separated in time by $\tau := t  - t'$. Both $t$ and $t'$ are taken to be larger than $r$ to avoid the transients being emitted from the point $|\bfx| = t = 0$, but are otherwise arbitrary. 

The correlation function \pref{Wttinv} evaluated at such a configuration reduces to
\bea \label{Wtildetaur}
\widetilde{\cW}(\tau,r)  & = &  -  \frac{1}{4 \pi^2 (\tau - i \delta)^2}  + \frac{ 2 \epsilon^2}{\tilde{g}^2 r^2} \bigg[ I_{-}(\tau + 2 r, c) - I_{-}(\tau, c) - I_{+}( \tau , c ) + I_{+}( \tau-  2r , c) \bigg] \\
& \ & \qquad + \frac{\epsilon}{8 \pi^2 r^2} \bigg[ \frac{1}{\tau -2 r - i \delta } - \frac{1}{\tau + 2 r - i \delta} \bigg] -  \frac{32 \pi^2 \epsilon^4}{ \tilde{g}^4 r^2} \bigg[ I_{-}( \tau , c ) + I_{+}( \tau , c )  \bigg] - \frac{\epsilon^2}{4 \pi^2 r^2 ( \tau - i \delta )^2} \nn \\
& \ &  \qquad \qquad \qquad +  \frac{2\epsilon^2}{\tilde{g}^2 r^2} \bigg[ \Phi\bigg( e^{ - \tfrac{2\pi (\tau - i \delta)}{\beta}}, 1 , \frac{c\beta}{2\pi} \bigg) + \Phi\bigg( e^{ + \tfrac{2\pi (\tau - i \delta)}{\beta}}, 1 , \frac{c\beta}{2\pi} \bigg)  - \frac{2\pi}{c\beta} \bigg] \nn
\eea
with the functions $\Phi$ and $I_{\mp}(\tau,c)$ as defined in the main text --- {\it c.f.}~eqs.~\pref{IpmDef} and \pref{PhiDef} --- and the limit $\delta \to 0^{+}$ understood to be taken at the end. The couplings are contained within the parameter 
\be
 c  =  \frac{16\pi^2 \epsilon}{\tilde{g}^2} \left( 1 + \frac{\lambda}{4\pi \epsilon}\right)\,,
\ee
and so $c \to \infty$ is the perturbative $\tilde g^2 \to 0$ limit.

With these variables the coincident limit is $\tau \to 0$; a limit controlled by the $c \tau \ll 1$ limit of $I_\pm$ and by the $\tau / \beta \ll 1$ of $\Phi$. We start first with $\Phi$, for which the integral representation \pref{PhiIntRep} is more useful than is the series definition of \pref{PhiDef} because $\tau/\beta \ll 1$ requires the behaviour of $\Phi(z,s,a)$ as $|z| \to 1$. Straightforward evaluation gives the asymptotic form 
\bea
& \ & \Phi\bigg( e^{ - \tfrac{2\pi (\tau - i \delta)}{\beta}}, 1 , \frac{c\beta}{2\pi} \bigg) + \Phi\bigg( e^{ + \tfrac{2\pi (\tau - i \delta)}{\beta}}, 1 , \frac{c\beta}{2\pi} \bigg)  - \frac{2\pi}{c\beta} \\
& \ & \qquad \qquad \qquad \simeq \ - \log\left( \sfrac{2\pi}{\beta} [ \tau - i \delta ] \right) - \log\left( - \sfrac{2\pi}{\beta} [ \tau - i \delta ] \right) - 2 \gamma - 2 \psi^{(0)}\left(\sfrac{c\beta}{2\pi}\right) - \frac{2\pi}{c\beta}  \ \ + \ \cO\left( \sfrac{\tau}{\beta} \right) \ , \nn
\eea
where $\gamma$ is the Euler-Mascheroni constant and $\psi^{(0)}(z) = \Gamma'(z)/\Gamma(z)$ is the digamma function (as defined in the main text). 

The $c \tau \ll 1$ limit for $I_\pm$ is similarly found using the series expansion (that applies for $z \in \mathbb{C}$ with $|\mathrm{Arg}(z)| < \pi$ and so not directly on the branch cut) 
\be
E_{1}(z) \simeq - \gamma - \log(z) - \sum_{k=1}^{\infty} \frac{(-z)^k}{k \cdot k!}
\ee
which is a convergent sum for any $z \in \mathbb{C}$ but is particularly useful when $|z| \ll 1$. This means that for $|c \tau | \ll 1$ we have 
\be
I_{\mp}(\tau, c) \ \simeq \ - \gamma - \log\big[ c (\tau - i \delta) \big] + \cO(c \tau) \qquad \qquad |c \tau| \ll 1 \ ,
\ee
and so the combinations appearing in the Wightman function are given by
\be
I_{-}( \tau , c ) + I_{+}( \tau , c ) \ \simeq \  - \log\left( c [ \tau - i \delta ] \right) - \log\left( - c[ \tau - i \delta ] \right) - 2 \gamma \ \ + \ \cO\left( c \tau \right) \ ,
\ee
as well as
\bea
I_{-}( \tau + 2 r , c ) + I_{+}( \tau - 2 r , c ) & \simeq & e^{2 c r} E_{1}\big[ 2 c (r - i \delta ) \big]  +  e^{2 c r} E_{1}\big[ 2 c (r + i \delta ) \big]  + \ \cO\left( c \tau \right) \nn \\
& \simeq & 2 e^{2 c r} E_{1}\big( 2 c r \big)  \ \ + \ \cO\left( c \tau \right) 
\eea
where in the last line we can safely take $\delta \to 0^{+}$. The final ingredient notes that for $\tau \ll r$ we have
\be
- \frac{1}{\tau + 2 r - i \delta} + \frac{1}{\tau -2 r - i \delta } \ \simeq \ - \frac{1}{r} \bigg[ \; 1 \; + \; \cO\left( \sfrac{\tau^2}{r^2} \right) \; \bigg] \ .
\ee

Using these expression in eq.~\pref{Wtildetaur} for the correlator $\widetilde{W}(\tau,r)$ and grouping terms reveals the coincident behaviour
\bea \label{nothadamard}
\widetilde{\cW}(\tau,t)  & \simeq &  - \; \frac{ 1 + ({\epsilon^2}/{r^2}) }{4 \pi^2 (\tau - i \delta)^2} + \frac{32 \pi^2 \epsilon^4}{ \tilde{g}^4 r^2} \bigg[ \log\left( c [ \tau - i \delta ] \right) + \log\left( - c[ \tau - i \delta ] \right) \bigg]  \\
& \ & \qquad - \frac{\epsilon}{8 \pi^2 r^3} + \frac{ 4 \epsilon^2}{\tilde{g}^2 r^2} \bigg[ \log\left( \sfrac{c\beta}{2\pi} \right)  - \psi^{(0)}\left(\sfrac{c\beta}{2\pi}\right) - \frac{\pi}{c\beta} +  e^{2 c r} E_{1}\big( 2 c r \big) \bigg] + \frac{64 \pi^2 \gamma \epsilon^4}{ \tilde{g}^4 r^2}  \nn
\eea
up to terms that vanish as $\tau \to 0$. 

This result is to be compared with the general Hadamard property \cite{Hadamard} for Wightman functions in arbitrary curved spacetimes, which states that the light-like (and coincident) limit is given by
\be  \label{hadamard}
\langle \Omega | \phi(x) \phi(x') | \Omega \rangle \ \simeq \ \frac{1}{8\pi^2} \bigg\{ \frac{\Delta^{1/2}(x,x')}{\sigma(x,x')} + V(x,x') \log\left| \frac{ \sigma(x,x') }{L^2} \right| + W_{\Omega}(x,x') \bigg\}
\ee
where $\sigma(x,x') = \frac{1}{2} \Delta s^2(x,x')$ is half the square of the geodesic separation between $x$ and $x'$, $L$ is a reference length scale introduced on dimensional grounds and the $i \delta$-prescription is omitted for brevity.\footnote{Including it would mean replacing $\sigma \to \sigma_{\delta}$ in the above formula, where $\sigma_{\delta}(x,x') := \sigma(x,x') + 2 i \delta [ \cT(x) - \cT(x') ] + \delta^2$ for any future-increasing function of time $\mathcal{T}$.} The power of this expression is lies in the fact that the functions $\Delta$, $V$ and $W_{\Omega}$ are all regular as $x \to x'$, with $\Delta$ (the van Vleck determinant) and $V$ being universal functions only of the local spacetime geometry (and independent of the state $|\Omega\rangle$). In particular $\Delta(x,x) = 1$ and $V(x,x) = 0$ for massless fields in a flat geometry. The robustness of this form relies on decoupling of scales, since quantum fluctuations of sufficiently short scales `forget' that they actually live in a curved spacetime.

For the situation at hand, with time-like separation $\Delta s^2 = - \tau^2$, Hadamard form becomes
\be 
\widetilde{\cW}_{\rm Had}(\tau,r) \simeq - \; \frac{ 1 }{4 \pi^2 (\tau - i \delta)^2} +  \hbox{(finite)} \,,
\ee
and so expression \pref{nothadamard} is {\it not} Hadamard, but becomes so in the limit $\epsilon/r \to 0$. Because $\epsilon$ is a UV scale associated with near-hotspot resolution in the effective theory for which it cannot be resolved from a point, \pref{nothadamard} correctly expresses how vacuum fluctuations of long-wavelength modes behave as they would in the absence of a hotspot but only do so if the coincident point is itself far from the hotspot. A coincident limit taken microscopically close to the hotspot in general can (and does) differ from a vanilla vacuum form, precisely because it is close enough to the hotspot for UV degrees of freedom to become relevant.

\subsection{Perturbative and large-separation limit of $\mathscr{S}$}
\label{App:curlySpert}

In this section we provide an asymptotic expression for $\mathscr{S}$ that applies both in the limit of large separations and in the perturbative limit where $\tilde g \to 0$.  
These limits are related because for $\mathscr{S}$ they both correspond to taking
\be
c \tau  =  \frac{16 \pi^2 \epsilon \tau}{\tilde{g}^2} \left(1  + \frac{\lambda}{4 \pi \epsilon} \right)   \gg 1 
\ee
in the functions $I_\pm(\tau,c)$. 

To find the asymptotic form in this limit we start with the following large-argument expression for the function $E_{1}(z)$, asymptotes to the series 
\be
E_{1}(z)\ \simeq \ e^{ - z} \bigg[ \frac{1}{z} - \frac{1}{z^2} + \cO(z^{-3})  \bigg] \qquad \qquad \mathrm{for}\ |z| \gg 1 \,.
\ee
Used in \pref{IpmDef} this implies that the functions $I_{\mp}(\tau, c )$ have the following asymptotic form for $| c \tau | \gg 1$
\be
I_{\mp}(\tau, c) \ \simeq \ \pm \frac{1}{c(\tau - i \delta)} - \frac{1}{c^2(\tau - i \delta)^2} + \cO \left(|c\tau|^{-3} \right) \qquad \qquad \mathrm{for}\  |c \tau| \gg 1 \ .
\ee
With these expressions the temperature-independent part of the Wightman function,  $\mathscr{S}$, becomes --- after dropping $\cO(|c\tau|^{-3})$ contributions, grouping terms and using $c = ( 16 \pi^2 \epsilon + 4 \pi \lambda ) / \tilde{g}^2$, 
\bea
\mathscr{S}(t,\bfx ; t', \bfx') & \simeq & \frac{1}{4 \pi^2 \left[ - (t - t' - i \delta)^2 + |\bfx - \bfx'|^2 \right]} \\
& \ & \qquad + \frac{ 1}{16 \pi^3 |\bfx| |\bfx'|} \cdot  \frac{\lambda}{1 + \frac{\lambda}{4 \pi \epsilon} } \cdot \frac{|\bfx| + |\bfx'|}{(t-t' - i \delta)^2 - (|\bfx|+|\bfx'|)^2} \nn \\
& \ & \qquad + \frac{\tilde{g}^2}{32 \pi^4 \big(1 + \frac{\lambda}{4 \pi \epsilon}\big)^2} \bigg[   - \frac{1}{|\bfx|} \sfrac{t-t'-|\bfx|}{\big[ (t - t' - |\bfx| - i \delta)^2 - |\bfx'|^2 \big]^2}  + \frac{1}{|\bfx'|} \sfrac{t-t'+|\bfx'|}{\big[ (t - t' + |\bfx'| - i \delta)^2 - |\bfx|^2 \big]^2}\bigg] \nn \\
& \  & \qquad - \frac{1}{ 64 \pi^4 |\bfx| |\bfx'|} \cdot \frac{\lambda^2}{\big( 1 + \frac{\lambda}{4 \pi \epsilon} \big)^2} \cdot \frac{1}{( t - t' -|\bfx| + |\bfx'|  - i \delta )^2 }   \ . \nn
\eea
For perturbatively small $\lambda$ --- {\it i.e.}~when $\lambda/(4\pi\epsilon) \ll 1$ ---  the leading part of this expression becomes 
\bea
\mathscr{S}(t,\bfx ; t', \bfx') & \simeq & \frac{1}{4 \pi^2 \left[ - (t - t' - i \delta)^2 + |\bfx - \bfx'|^2 \right]} \\
& \ & \qquad + \frac{\lambda}{16 \pi^3 |\bfx| |\bfx'|} \cdot \frac{|\bfx| + |\bfx'|}{(t-t' - i \delta)^2 - (|\bfx|+|\bfx'|)^2} \nn \\
& \ & \qquad + \frac{\tilde{g}^2}{32 \pi^4} \bigg[   - \frac{1}{|\bfx|} \frac{t-t'-|\bfx|}{\big[ (t - t' - |\bfx| - i \delta)^2 - |\bfx'|^2 \big]^2}  + \frac{1}{|\bfx'|} \frac{t-t'+|\bfx'|}{\big[ (t - t' + |\bfx'| - i \delta)^2 - |\bfx|^2 \big]^2}\bigg] \nn 
\eea
and so agrees with the temperature-independent part of expression \pref{pertcorr} used in the main text.

\subsection{Perturbative Limit of $\mathscr{E}_\beta$}
\label{App:curlyEpert}

The large-$\tau$ and small $\tilde g^2$ limits are not equivalent for the temperature-dependent part of the correlator. The perturbative limit of $\mathscr{E}_\beta$ corresponds to the regime
\be
\frac{c\beta}{2\pi} \gg 1\,,
\ee
for which the relevant large-$a$ asymptotic representation of the Lerch transcendent is (for $a > 0$)
\be
\Phi(z,s,a) \simeq \frac{1}{1-z} \left( \frac{1}{a^s} \right)+ \sum_{n=1}^{N-1} \frac{(-1)^n \Gamma(s+n)}{n!\; \Gamma(s)} \cdot  \frac{\mathrm{Li}_{-n}(z)}{a^{s+n}} + \cO(a^{-s-N}) \qquad \quad \mathrm{for\ } a \gg 1
\ee
for fixed $s \in \mathbb{C}$ and fixed $z \in \mathbb{C}  \setminus [ 1, \infty )$, where 
\be
   \mathrm{Li}_{-n}(z) = \left( z \partial_{z} \right)^n \frac{z}{1-z}
\ee
are polylogarithm functions of negative integer order, of which the required particular cases are
\be
  \mathrm{Li}_{-1}(z) = \frac{z}{(1-z)^2}\,, \quad
  \mathrm{Li}_{-2}(z) = \frac{z+ z^2}{(1-z)^3}
\ee

Collecting terms, we find that for $c \beta \gg 2\pi$ we have 
\bea
\Phi\left( z, 1 , \sfrac{c\beta}{2\pi} \right) & \simeq &  \frac{1}{1 - z}\left( \frac{2\pi}{c\beta}\right) - \frac{z}{(1-z)^2} \bigg( \frac{2\pi}{c\beta} \bigg)^2 + \sfrac{z+ z^2}{(1-z)^3}  \bigg( \frac{2\pi}{c\beta} \bigg)^3   + \cO\left[ (c\beta)^{-4} \right] \nn\\
 \Phi\left( \sfrac{1}{z}, 1 , \sfrac{c\beta}{2\pi} \right) & \simeq &  \bigg[ 1 - \frac{1}{1 - z} \bigg] \frac{2\pi}{c\beta} - \frac{z}{(1-z)^2} \bigg( \frac{2\pi}{c\beta} \bigg)^2 - \sfrac{z+ z^2}{(1-z)^3} \bigg( \frac{2\pi}{c\beta} \bigg)^3  + \cO\left[ (c\beta)^{-4} \right]
\eea
and so
\be
\frac{1}{2}\Phi\left( z, 1 , \sfrac{c\beta}{2\pi} \right) +\frac{1}{2}\Phi\left( \sfrac{1}{z}, 1 , \sfrac{c\beta}{2\pi} \right)    \simeq   \frac{\pi}{c\beta} - \frac{ z}{(1-z)^2} \bigg( \frac{2\pi}{c\beta} \bigg)^2 + \cO\left[ (c\beta)^{-4} \right] \,. 
\ee
This allows the perturbative expression for $\mathscr{E}_\beta$ to be written
\bea
\mathscr{E}_\beta(t,\bfx ; t', \bfx') &\simeq &  -  \frac{\tilde{g}^2}{64 \pi^2 \beta^2 |\bfx| |\bfx'|\left( 1 + \frac{\lambda}{4 \pi \epsilon} \right)^2} \, \mathrm{csch}^2\left[ \frac{\pi [t - t' - |\bfx| + |\bfx'| - i \delta]}{\beta} \right] + \ldots 
\eea
the first term of which exactly captures the temperature-dependent terms in the perturbative result quoted in formula \pref{pertcorr} in the main text. When evaluated with $|\bfx | = |\bfx'|$ these reveal the exponential fall-off described in the main text when $\tau \gg \beta$.

\section{Qubit Integrals}
\label{App:UDWintegrals}

This appendix evaluates the integrals appearing in the expressions for $\cC$, $\cD$ and $\cR$ in the main text. 

\subsection{Exact integrals}

The Wightman function has here the form given in (\ref{Wttinv})
\be
\widetilde{\cW}(\tau)  : = \mathscr{S}(\tau, \bfx_{\ssQ} ; 0, \bfx_{\ssQ} ) |_{\lambda = 0} + \mathscr{E}_{\beta}(\tau, \bfx_{\ssQ} ; 0, \bfx_{\ssQ} ) |_{\lambda =0}
\ee
with the functions $\mathscr{S}$ and $\mathscr{E}$ defined in \pref{curlySanswer} and \pref{curlyEanswer}. To simplify the calculation of the required integrals we split apart the Wightman function into three pieces such that
\be
\widetilde{\cW}(\tau) : = \ \widetilde{\cW}_1(\tau) + \widetilde{\cW}_2(\tau) + \widetilde{\cW}_3(\tau)
\ee
where we define
\bea
\widetilde{\cW}_1(\tau) : =  - \sfrac{1}{4\pi^2 (\tau - i \delta)^2} \quad , \qquad \widetilde{\cW}_2(\tau) := \mathscr{S}(\tau, \bfx_{\ssQ} ; 0, \bfx_{\ssQ} ) |_{\lambda = 0} -\widetilde{\cW}_1(\tau) \\
\mathrm{and} \qquad \widetilde{\cW}_3(\tau) : = \mathscr{E}_{\beta}(\tau, \bfx_{\ssQ} ; 0, \bfx_{\ssQ} ) |_{\lambda = 0 }\ . \qquad \qquad \qquad \qquad \qquad \nn
\eea
It turns out that the momentum space representation of the above functions are most useful here, where
\be
\widetilde{\cW}_1(\tau) = \frac{1}{4 \pi^2} \int_0^\infty \exd p\; p e^{- i p \tau} \ , \label{wideW1mom}
\ee
and $\widetilde{\cW}_2(\tau)$ and $\widetilde{\cW}_3(\tau)$ can be written in momentum space as (see \cite{Hotspot})
\bea
\widetilde{\cW}_2(\tau) & = & \frac{\epsilon}{4 \pi^2 |\bfx_{\ssQ}|^2}  \int_0^\infty \exd p\; e^{- i p \tau } \bigg( \sin( |\bfx_{\ssQ}| p ) \cdot 2  \mathrm{Re} \bigg[ e^{- i p |\bfx_{\ssQ}|} \frac{- i p}{ c_0 + i p } \bigg] + \frac{\epsilon p^3}{c_0^2 + p^2} \bigg)  \label{wideW2mom} 
\eea
and
\be
\widetilde{\cW}_3(\tau) =  \frac{4\epsilon^2}{\tilde{g}^2 |\bfx_{\ssQ}|^2} \int_0^\infty \exd p\;  \frac{p}{p^2 + c_0^2} \bigg[ \; e^{- i p \tau} + \frac{2 \cos( p \tau )}{e^{ \beta p }  - 1 } \; \bigg] \ , \label{wideW3mom}
\ee
which we write in terms of the parameter 
\be
c_0 \ := \ c \; |_{\lambda = 0} \ = \ \frac{16 \pi^2 \epsilon}{\tilde{g}^2} 
\ee
as in the main text. Notice that each of these functions can be written in the form
\be
\widetilde{W}_j(\tau) \ = \ \int_0^\infty \exd p\; \bigg[ \cos( p \tau) F_{j}(p) - i \sin(p \tau) G_{j}(p) \bigg] \ , 
\ee
where $F_{j}$ and $G_{j}$ are the real-valued functions
\bea
F_{1}(p) &=& G_{1}(p) \ : = \  \frac{p}{4 \pi^2} \nn \\
F_{2}(p) &=& G_{2}(p) \ : = \ \frac{\epsilon}{4 \pi^2 |\bfx_{\ssQ}|^2} \bigg( \sin( |\bfx_{\ssQ}| p ) \cdot 2  \mathrm{Re} \bigg[ e^{- i p |\bfx_{\ssQ}|} \frac{- i p}{ c_0 + i p } \bigg] + \frac{\epsilon p^3}{c_0^2 + p^2} \bigg) \\
F_{3}(p) & := & \frac{4\epsilon^2}{\tilde{g}^2 |\bfx_{\ssQ}|^2} \frac{p}{p^2 + c_0^2} \coth\left( \frac{\beta p}{2} \right) \qquad \mathrm{and} \qquad G_{3}(p) := \frac{4\epsilon^2}{\tilde{g}^2 |\bfx_{\ssQ}|^2}  \frac{p}{p^2 + c_0^2}  \ . \nn
\eea
We now compute the integrals for $j \in \{ 1, 2, 3 \}$,
\bea
\mathcal{C}_{j} & := & 2 \int_0^\infty \exd s\; \mathrm{Re}[\widetilde{\cW}_{j}(s)] \cos(\omega s) \nn \\
\mathcal{S}_{j} & := & 2 \int_0^\infty \exd s\; \mathrm{Im}[\widetilde{\cW}_{j}(s)] \sin(\omega s) \\
\mathcal{D}_{j} & := & 2 \int_0^\infty \exd s\; \mathrm{Re}[\widetilde{\cW}_{j}(s)] \sin(\omega s) \nn
\eea
where $\cC = \sum_{j} \cC_{j}$, $\cR = \sum_{j} ( \cC_{j} + \cS_{j} )$ as well as $\cD = \sum_{j} \cD_{j}$ give the functions \pref{Cqubitdef} and \pref{Rqubitdef} defined in the main text. Recall that we assume $\omega > 0$. To compute the functions $\cC_{j}$ we find
\bea
\cC_{j} & = & 2 \int_0^\infty \exd s\; \cos(\omega s)  \int_0^\infty \exd p\; \cos(p s) F_{j}(p) \\
& = & \int_0^\infty \exd p\; F_{j}(p) \int_0^\infty \exd s\; \bigg( \cos( [ p - \omega ] s) + \cos( [ p - \omega ] s) \bigg) \nn \\
& = & \pi \int_0^\infty \exd p\; F_{j}(p) \bigg( \delta( p - \omega ) + \delta( p - \omega ) \bigg) \nn \\
& = & \pi F_{j}(\omega) \ . \nn
\eea
Similarly for the functions $\cS_{j}$ we find 
\bea
\cS_{j} & = &  - 2 \int_0^\infty \exd s \; \sin(\omega s)\int_0^\infty \exd p\; \sin(p s) G_{j}(p)  \\
& = & - \int_0^\infty \exd p\; G_{j}(p) \int_0^\infty \exd s\; \bigg( \cos( [ p - \omega ] s) - \cos( [ p - \omega ] s) \bigg) \nn \\
& = & - \pi \int_0^\infty \exd p\; G_{j}(p) \bigg( \delta( p - \omega ) - \delta( p - \omega ) \bigg) \nn \\
& = & - \pi G_{j}(\omega) \ . \nn
\eea
Summing the above functions to get $\cC$ gives the quoted answer in \pref{cCIntegralResult}
\be
\cC = \sum_{j=1}^{3} \cC_{j} = \pi \sum_{j=1}^{3} F_{j}(\omega) = \frac{\omega}{4 \pi} \bigg( 1 + \sfrac{\tilde{g}^2\big[ \frac{\tilde{g}^2\omega^2}{16 \pi^2} + \cos\big( 2 \omega |\bfx_{\ssQ}| \big) - 1 - \dfrac{\tilde{g}^2 \omega}{16 \pi^2 \epsilon}  \sin\big( 2 \omega |\bfx_{\ssQ}|  \big) + \mathrm{coth}\big( \frac{\beta \omega}{2} \big)  \big] }{16 \pi^2 |\bfx_{\ssQ}|^2 \big[ ( \frac{\tilde{g}^2 \omega}{16 \pi^2 \epsilon} )^2 + 1 ] }  \bigg)
\ee
after some simplification, as well as the quoted answer \pref{cRIntegralResult}
\be
\cR = \sum_{j=1}^{3} ( \cC_{j} + \cS_{j} )  = \pi \big[ F_{3}(\omega) - G_{3}(\omega) \big] = \frac{\tilde{g}^2 \omega }{32 \pi^3 |\bfx_{\ssQ}|^2 \big[  ( \frac{\tilde{g}^2 \omega}{16 \pi^2 \epsilon} )^2 + 1 \big] \left( e^{\beta \omega} - 1 \right)} \ . \qquad
\ee
For the functions $\cD_j$ we must compute
\bea
\cD_{j} & = & 2 \int_0^\infty \exd s\; \sin(\omega s) \int_0^\infty \exd p\; \cos(p s) F_{j}(p) \\
& = & - \int_0^\infty \exd p\; F_{j}(p) \int_0^\infty \exd p\; \bigg( \sin\big( [ p - \omega ] s \big) - \sin\big( [ p + \omega ] s \big) \bigg) \nn \\
& = & - \mathcal{PV} \int_0^\infty \exd p\; F_{j}(p) \bigg( \frac{1}{p - \omega} - \frac{1}{p + \omega} \bigg) \nn
\eea
where the integral over the singularity at $p = \omega$ is a Cauchy Principal value (which follows from taking the imaginary part of $\int_{-\infty}^{\infty} \exd x\; e^{- i y x }\Theta(x) = \frac{- i }{y - i \delta}$). The function $\cD = \sum_{j} \cD_{j}$ also turns out to be ultraviolet divergent, and so we impose a momentum cutoff $\Lambda$ on the integrals here so that 
\bea
\cD_{j} & = & 2 \omega \cdot \mathcal{PV} \int_0^\Lambda \exd p\; \frac{F_{j}(p)}{\omega^2 - p^2}\ . 
\eea
First we compute $\cD_{1}$ to find
\bea
\cD_{1} & = & \frac{\omega}{2\pi^2} \cdot \mathcal{PV} \int_0^\Lambda \exd p\; \frac{p}{\omega^2 - p^2} \ = \ \frac{\omega}{2\pi^2} \log\left( \sfrac{\omega}{\sqrt{\Lambda^2-\omega^2}} \right) \ \simeq \ \frac{\omega}{2\pi^2} \bigg[ \log\left( \frac{\omega}{\Lambda} \right) + \cO\bigg( \sfrac{\omega^2}{\Lambda^2} \bigg)   \bigg]  \ .  
\eea
where we have taken the limit $\Lambda \gg \omega$ in the last equality. For the next function $\cD_{2}$ we have 
\bea
\cD_{2} & = & \frac{\omega \epsilon}{2 \pi^2 |\bfx_{\ssQ}|^2} \cdot \mathcal{PV} \int_0^\Lambda \exd p\; \frac{1}{\omega^2 - p^2} \bigg( \sin( |\bfx_{\ssQ}| p ) \cdot 2  \mathrm{Re} \bigg[ e^{- i p |\bfx_{\ssQ}|} \frac{- i p}{ c_0 + i p } \bigg] + \frac{\epsilon p^3}{c_0^2 + p^2} \bigg) \\
& = & - \frac{\omega \epsilon }{2 \pi^2 |\bfx_{\ssQ}|^2} \cdot \mathcal{PV} \int_0^\Lambda \frac{\exd p}{\omega^2 - p^2} \cdot \frac{p^2 \sin\big(2 p |\bfx_{\ssQ}|\big) }{p^2 + c_0^2} - \frac{\omega \epsilon c_0 }{2 \pi^2 |\bfx_{\ssQ}|^2} \cdot \mathcal{PV} \int_0^\Lambda \frac{\exd p}{\omega^2 - p^2} \cdot \frac{p \big[ 1 -  \cos\big(2 p |\bfx_{\ssQ}|\big) \big] }{p^2 + c_0^2} \nn \\
& \ & \qquad \qquad \qquad + \frac{\omega \epsilon^2 }{2 \pi^2 |\bfx_{\ssQ}|^2} \cdot \mathcal{PV} \int_0^\Lambda \frac{\exd p}{\omega^2 - p^2} \cdot \frac{p^3}{p^2 + c_0^2} \ . \nn
\eea
This can be rewritten as (recall that $c_0 = 16 \pi^2 \epsilon/\tilde{g}^2$)
\bea
\cD_{2} & = & \frac{\omega \epsilon^2 }{2 \pi^2 |\bfx_{\ssQ}|^2} \cdot \mathcal{PV} \int_0^\Lambda \frac{\exd p}{\omega^2 - p^2} \cdot \frac{p^3}{p^2 + c_0^2} \ \  - \frac{\omega \epsilon }{2 \pi^2 |\bfx_{\ssQ}|^2} \cdot \mathcal{PV} \int_0^\Lambda \frac{\exd p}{\omega^2 - p^2} \cdot \frac{p^2 \sin\big(2 p |\bfx_{\ssQ}|\big) }{p^2 + c_0^2} \qquad \qquad  \\
& \ & \qquad + \frac{8 \omega \epsilon^2}{\tilde{g}^2 |\bfx_{\ssQ}|^2} \cdot \mathcal{PV} \int_0^\Lambda \frac{\exd p}{\omega^2 - p^2} \cdot \frac{p \cos\big(2 p |\bfx_{\ssQ}|\big) }{p^2 + c_0^2} - \frac{8 \omega \epsilon^2}{\tilde{g}^2 |\bfx_{\ssQ}|^2} \cdot \mathcal{PV} \int_0^\Lambda \frac{\exd p}{\omega^2 - p^2} \cdot \frac{p}{p^2 + c_0^2} \nn
\eea
and for $\cD_{3}$ we have 
\bea
\cD_{3} & = & \frac{8 \omega \epsilon^2}{\tilde{g}^2 |\bfx_{\ssQ}|^2} \cdot \mathcal{PV} \int_0^\Lambda \exd p\; \frac{1}{\omega^2 - p^2} \cdot \frac{p}{p^2 + c_0^2} \coth\left( \frac{\beta p}{2} \right) \ .
\eea
Summing $\cD_{2} + \cD_{3}$ (and grouping the last term of $\cD_2$ with $\cD_3$ by using $\coth( x/2 ) - 1 = 2(e^x - 1)^{-1}$), we find that we need to compute four separate integrals
\be
\cD_{2} + \cD_{3} \ = \ \frac{\omega \epsilon^2 }{2 \pi^2 |\bfx_{\ssQ}|^2} I^{(\mathrm{div})}_1 \ - \  \frac{\omega \epsilon }{2 \pi^2 |\bfx_{\ssQ}|^2} I_2  \ + \ \frac{8 \omega \epsilon^2}{\tilde{g}^2 |\bfx_{\ssQ}|^2 } \big[ \; I_3 \ + \ 2 I_4 \; \big]
\ee
with the four integrals defined by
\bea \label{Idiv1}
I^{(\mathrm{div})}_1 & := &  \mathcal{PV} \int_0^\Lambda \frac{\exd p}{\omega^2 - p^2} \cdot \frac{p^3}{p^2 + c_0^2}  \\
I_{2} & : = & \mathcal{PV} \int_0^\infty \frac{\exd p}{\omega^2 - p^2} \cdot \frac{p^2 \sin\big(2 p |\bfx_{\ssQ}|\big) }{p^2 + c_0^2} \nn \\
I_{3} & := & \mathcal{PV} \int_0^\infty \frac{\exd p}{\omega^2 - p^2} \cdot \frac{p \cos\big(2 p |\bfx_{\ssQ}|\big) }{p^2 + c_0^2} \nn \\
I_{4} & := & \mathcal{PV} \int_0^\infty \exd p\; \frac{1}{\omega^2 - p^2} \cdot \frac{p}{p^2 + c_0^2} \cdot \frac{1}{e^{\beta p} - 1} \ . \nn
\eea
Only the first integral is divergent for large $\Lambda$ and happens to be elementary, where
\bea
I^{(\mathrm{div})}_1 & = & \mathcal{PV} \int_0^\Lambda \exd p\; \frac{ p}{\omega^2 - p^2} \ - \ c_0^2 \cdot \mathcal{PV} \int_0^\Lambda \frac{\exd p\;}{\omega^2 - p^2} \cdot \frac{p}{p^2 + c_0^2} \label{divI1split} \\
& = & \log\left( \sfrac{\omega}{\sqrt{ \Lambda^2 - \omega^2 } } \right) - \frac{c_0^2}{2(c_0^2+ \omega^2)} \log \left( \frac{(\Lambda/c_0)^2 + 1}{(\Lambda/\omega)^2 + 1} \right) \ . \nn
\eea
Assuming that $\Lambda \gg \omega$ (as above) yields
\bea
I^{(\mathrm{div})}_1 & \simeq & \log\left( \frac{\omega}{ \Lambda } \right) - \frac{c_0^2}{2(c_0^2+ \omega^2)} \log \left( \frac{(\Lambda/c_0)^2 + 1}{(\Lambda/\omega)^2} \right)
\eea
From here we notice that the second term is actually UV-finite if one assumes $\Lambda \gg c_0$ (as can also be seen by power-counting the second integral in \pref{divI1split}) where\footnote{Note that in the perturbative limit, one needs instead $\Lambda \ll c_0$ to be true (see \cite{Hotspot}). In this limit it is easy to see that $I^{(\mathrm{div})}_1 \simeq 0$, which shows how the cutoff dependence matches the Hadamard structure of the perturbative limit of $\widetilde{\cW}(\tau)$ --- {\it ie.} the divergent part of $\cD$ has no $|\bfx_{\ssQ}|$-dependence in the perturbative limit.}
\be
I^{(\mathrm{div})}_1 \ \simeq \ \log\left( \dfrac{\omega}{\Lambda} \right) - \dfrac{c_0^2}{c_0^2 + \omega^2} \log\left( \frac{\omega}{c_0} \right)\ .
\ee
It turns out that the remaining three integral $I_2$, $I_3$ and $I_4$ defined in \pref{Idiv1} are all UV finite, and so they can safely have their upper limits taken to $\simeq \infty$. To compute $I_2$ we write
\be
I_2 =  \frac{\omega}{2( c_0^2 +\omega^2 )} \bigg[ \int_0^{\infty} \exd p\; \sfrac{\sin\big(2|\bfx_\ssQ| p\big)}{p+\omega}  - \mathcal{PV} \int_0^{\infty} \exd p\; \sfrac{\sin\big(2 |\bfx_\ssQ| p \big)}{p-\omega} \bigg] - \frac{c_0^2}{c_0^2 + \omega^2} \int_0^{\infty} \exd p\; \sfrac{\sin\big(2|\bfx_\ssQ| p\big)}{p^2+c_0^2} 
\ee
Note that only one of these integrals is a principal value integral after the partial fraction decomposition. Using formulae (3.722.1), (3.722.5) and (3.723.1) from \cite{grad} the above is easily seen to evaluate to
\bea
I_2 & = & \frac{\omega}{c_0^2 +\omega^2 } \bigg[ \mathrm{Ci}\big(2 |\bfx_{\ssQ}| \omega\big) \sin( 2 |\bfx_{\ssQ}| \omega ) - \mathrm{Si}\big(2 |\bfx_{\ssQ}| \omega\big) \cos( 2 |\bfx_{\ssQ}| \omega ) \bigg] \\
& \ & \qquad \qquad \qquad - \frac{c_0 e^{ - 2  c_0 |\bfx_{\ssQ}|} }{2 ( c_0^2 + \omega^2 ) } \; \mathrm{Ei}\big(2|\bfx_{\ssQ}|c_0\big) + \frac{c_0 e^{ 2 |\bfx_{\ssQ}| c_0} }{2 ( c_0^2 + \omega^2 ) } \; \mathrm{Ei}\big(-2|\bfx_{\ssQ}|c_0\big) \nn
\eea
where $\mathrm{Si}(z) := \int_0^z \exd t\; \frac{\sin(t)}{t}$ is the sine integral function, $\mathrm{Ci}(z) : = - \int_{z}^{\infty} \exd t\; \frac{\cos(t)}{t}$ is the cosine integral function and $\mathrm{Ei}(z) = - \int_{-z}^{\infty} \exd t\; \frac{e^{-t}}{t}$ is the exponential integral function.

In a very similar computation, we use formulae (3.722.3), (3.722.7) and (3.723.5) from \cite{grad} to compute $I_{3}$ where
\bea
I_3 & = & \frac{1}{c_0^2 +\omega^2 } \bigg[ \mathrm{Ci}\big(2 |\bfx_{\ssQ}| \omega\big) \cos( 2 |\bfx_{\ssQ}| \omega ) + \mathrm{Si}\big(2 |\bfx_{\ssQ}| \omega\big) \sin( 2 |\bfx_{\ssQ}| \omega ) \bigg] \\
& \ & \qquad \qquad \qquad - \frac{e^{ - 2 |\bfx_{\ssQ}| c_0} }{2 ( c_0^2 + \omega^2 ) } \; \mathrm{Ei}\big(2|\bfx_{\ssQ}|c_0\big) - \frac{e^{ 2 |\bfx_{\ssQ}| c_0} }{2 ( c_0^2 + \omega^2 ) } \; \mathrm{Ei}\big(-2|\bfx_{\ssQ}|c_0\big) \nn
\eea
And finally we compute $I_{4}$ where
\bea
I_{4} & = & \frac{1}{c_0^2 + \omega^2} \bigg[ \int_0^\infty \exd p\; \frac{p}{p^2 + c_0^2} \cdot \frac{1}{e^{\beta p} - 1} - \mathcal{PV} \int_0^\infty \exd p\; \frac{p}{p^2 - \omega^2} \cdot \frac{1}{e^{\beta p} - 1} \bigg]
\eea
To compute this integral we note the integral representation (see formula (5.9.15) in \cite{NIST}) of the digamma function, defined by $\psi^{(0)}(z) := \Gamma'(z) / \Gamma(z)$, where
\be
\psi^{(0)}(z) = \log(z) - \frac{1}{2z} - 2 \int_0^\infty \exd t\; \frac{t}{t^2 + z^2} \cdot \frac{1}{e^{2 \pi t}  - 1}
\ee
for any $z \in \mathbb{C}$ with $\mathrm{Re}[z]>0$. It is easily seen from this expression that 
\be
\int_0^\infty \exd p\; \frac{p}{p^2 + c_0^2} \cdot \frac{1}{e^{\beta p} - 1}  = \frac{1}{2} \log\left( \frac{\beta c_0}{2\pi} \right) - \frac{1}{2} \psi^{(0)}\left( \frac{\beta c_0}{2\pi} \right) - \frac{\pi}{2 \beta c_0} \ .
\ee
The other (principal value) integral also follows from the above integral representation --- fixing $z = \delta + i \omega$ and taking the limit $\delta \to 0^{+}$ (and then taking the real part of both sides of the expression) yields
\be
\mathcal{PV} \int_0^\infty \exd p\; \frac{p}{p^2 - \omega^2} \cdot \frac{1}{e^{\beta p} - 1} = \frac{1}{2} \log\left( \frac{\beta \omega}{2\pi} \right) + \frac{1}{2} \mathrm{Re}\left[ \psi^{(0)}\left( i \frac{\beta \omega}{2\pi} \right) \right] \ ,
\ee
giving
\bea
I_{4} = \frac{1}{2( c_0^2 + \omega^2)} \bigg[ \log\left( \frac{c_0}{\omega} \right) - \psi^{(0)}\left( \frac{\beta c_0}{2\pi} \right) - \mathrm{Re}\left[ \psi^{(0)}\left( i \frac{\beta \omega}{2\pi} \right) \right] - \frac{\pi}{\beta c_0} \bigg] \ .
\eea
Putting this all together into the sum $\cD = \sum_{j=1}\cD_{j}$ and simplifying (using $c_0 = 16\pi^2 \epsilon / \tilde{g}^2$ where necessary) leaves us with the function quoted in \pref{Danswer} in the main text:
\bea \label{App:Danswer}
\cD & = & \frac{\omega}{2\pi^2} \bigg[ 1 + \frac{\epsilon^2}{|\bfx_{\ssQ}|^2} \bigg]  \log\left( \frac{\omega}{\Lambda} \right) - \frac{\omega \epsilon^2}{2\pi^2 |\bfx_{\ssQ}|^2} \cdot \frac{1}{ (\omega/c_0)^2 + 1 } \log\left( \frac{\omega}{c_0} \right) \nn \\
& \ & \quad -  \frac{\tilde{g}^2\omega}{32 \pi^4 |\bfx_{\ssQ}|^2}  \cdot \frac{(\omega/c_0)}{(\omega/c_0)^2+ 1} \cdot \bigg[ \mathrm{Ci}\big(2 |\bfx_{\ssQ}| \omega\big) \sin( 2 |\bfx_{\ssQ}| \omega ) - \mathrm{Si}\big(2 |\bfx_{\ssQ}| \omega\big) \cos( 2 |\bfx_{\ssQ}| \omega ) \bigg]  \\
& \ & \quad  + \frac{\tilde{g}^2 \omega}{32 \pi^4 |\bfx_{\ssQ}|^2} \cdot \frac{1}{(\omega/c_0)^2 + 1} \cdot \bigg\{ \mathrm{Ci}\big(2 |\bfx_{\ssQ}| \omega\big) \cos( 2 |\bfx_{\ssQ}| \omega ) + \mathrm{Si}\big(2 |\bfx_{\ssQ}| \omega\big) \sin( 2 |\bfx_{\ssQ}| \omega )  \nn \\
& \ & \qquad \qquad \qquad \qquad - e^{ - 2 |\bfx_{\ssQ}| c_0} \mathrm{Ei}\big(2|\bfx_{\ssQ}|c_0\big)  - \log\left( \frac{\omega}{c_0} \right) - \psi^{(0)}\left( \frac{\beta c_0}{2\pi} \right) + \mathrm{Re}\left[ \psi^{(0)}\left( i \frac{\beta \omega}{2\pi} \right) \right] - \frac{\pi}{\beta c_0} \bigg\} \nn \ .
\eea

\section{Control over the Markovian approximation}
\label{App:MarkovControl}

In this Appendix we fill in the details of the identification of the region of parameter space in which the Markovian approximation applies. We assume the parameter regime \pref{parameters} --- or, more usefully, \pref{parameters2} --- required to obtain thermalization at the hotspot temperature, which imply
\be
\frac{\tilde{g} \omega}{4 \pi} , \frac{4\pi\epsilon}{\tilde{g}} \ll \frac{4 \pi |\bfx_{\ssQ}|}{\tilde{g}} \ll 1 \quad \text{and} \quad\omega |\bfx_{\ssQ}| \ll \frac{4\pi\epsilon}{\tilde{g}}  \ .
\ee

In this parameter regime, the function $\cC$ given in \pref{cCIntegralResult} has the approximate form
\be
\cC \ \simeq \ \frac{\omega}{4 \pi} \bigg[ 1 + \frac{\tilde{g}^2 \mathrm{coth}\big( \tfrac{\beta \omega}{2} \big)  }{16 \pi^2 |\bfx_{\ssQ}|^2 \left[  (\omega/c_0)^2 + 1 \right]} \bigg] \ ,
\ee
while its $\omega$-derivative becomes
\be \label{Cprime}
\frac{\exd \mathcal{C}}{\exd \omega} \ \simeq\  \frac{1}{4 \pi} \bigg[ 1 + \frac{\tilde{g}^2   }{16 \pi^2 |\bfx_{\ssQ}|^2} \bigg( \frac{  \mathrm{coth}\big( \tfrac{\beta \omega}{2} \big) - \frac{\beta \omega}{2} \mathrm{csch}^2 \big( \tfrac{\beta \omega}{2} \big)  }{ (\omega/c_0)^2 + 1 } - \frac{2 (\omega/c_0)^2 \coth\big( \tfrac{\beta \omega}{2} \big)}{\left[ (\omega/c_0)^2 +1 \right]} \bigg) \bigg] \ .
\ee
In the same parameter regime the function $\cD$ takes the approximate form\footnote{We use here $\mathrm{Ci}(z) \simeq \log(e^\gamma z) + \cO(z^2)$, $\mathrm{Si}(z) \simeq z + \cO(z^3)$ and $\mathrm{Ei}(\pm z) \simeq \log(e^{\gamma}z) + \cO(z)$ for $0< z \ll 1$. Note also that the combination $c_0 |\bfx_{\ssQ}| = \frac{16 \pi^2 \epsilon |\bfx_{\ssQ}|}{\tilde{g}^2} = \frac{4\pi \epsilon}{\tilde{g}} \cdot \frac{4 \pi |\bfx_{\ssQ}|}{\tilde{g}} \ll 1$ is small in the considered parameter regime.}
\bea
\cD & \simeq & \frac{\omega}{2\pi^2} \bigg[ \log\left( \frac{\omega}{\Lambda} \right) - \frac{\epsilon^2}{|\bfx_{\ssQ}|^2} \cdot \frac{1}{ (\omega/c_0)^2 + 1 } \cdot \log\big( \frac{\omega}{c_0} \big)  \\
& \ & \qquad \qquad \qquad \qquad + \frac{\tilde{g}^2}{16 \pi^2 |\bfx_{\ssQ}|^2} \cdot \frac{1}{(\omega/c_0)^2 + 1} \cdot \bigg\{  \mathrm{Re}\left[ \psi^{(0)}\left( i \frac{\beta \omega}{2\pi} \right) \right]  - \psi^{(0)}\left( \frac{\beta c_0}{2\pi} \right)- \frac{\pi}{\beta c_0} \bigg\} \bigg] \nn \ ,
\eea
and its $\omega$-derivative becomes
\bea \label{Dprime}
\frac{\exd \cD}{\exd \omega} & \simeq & \frac{1}{2\pi^2} \bigg[ \log\left( \sfrac{\omega}{\Lambda} e^1 \right) - \frac{\epsilon^2}{|\bfx_{\ssQ}|^2} \cdot \frac{2(\omega/c_0)^2 + \left[ 1 - (\omega/c_0)^2 \right]\log\big( \frac{\omega}{c_0} \big)}{ \left[ (\omega/c_0)^2 + 1 \right]^2 }  \\
&& \qquad \qquad + \sfrac{\tilde{g}^2}{16 \pi^2 |\bfx_{\ssQ}|^2} \cdot \sfrac{1}{(\omega/c_0)^2 + 1} \cdot \left\{ \sfrac{\beta \omega}{2\pi}  \mathrm{Im}\left[ \psi^{(1)}\left( i \sfrac{\beta \omega}{2\pi} \right) \right] \right. \nn \\
&& \qquad\qquad \qquad \qquad \qquad \qquad  \qquad \qquad  \left. + \sfrac{3(\omega/c_0)^2 +1}{(\omega/c_0)^2 + 1} \left(  \mathrm{Re}\left[ \psi^{(0)}\left( i \sfrac{\beta \omega}{2\pi} \right) \right] - \psi^{(0)}\left( \sfrac{\beta c_0}{2\pi} \right)- \sfrac{\pi}{\beta c_0} \right) \right\} \bigg] \nn \ .
\eea
where $\psi^{(1)}(z) := \frac{\mathrm{d}}{\mathrm{d}z} \psi^{(0)}(z) =  \frac{\mathrm{d}^2}{\mathrm{d}z^2} \log\Gamma(z)$.

\subsection{Non-degenerate limit}

We first identify the parameter range that satisfies the conditions $\left| {\lambda_{\ssQ}^2 \cC}/{\omega} \right| \ll 1$ and $\left| {\lambda_{\ssQ}^2 \cD}/{\omega} \right| \ll 1$ given in \pref{nondegen}, that were imposed when deriving (\ref{12sol1}) in the `non-degenerate' limit (which is done out of convenience rather than absolute necessity). 

To this end we record the following approximate forms for the expressions for ${\lambda_{\ssQ}^2 \cC}/{\omega}$ and ${\lambda_{\ssQ}^2 \cD}/{\omega}$,
\bea \label{NonDegenExplicit}
\frac{\lambda_{\ssQ}^2 \cC}{\omega} & \simeq & \frac{\lambda_{\ssQ}^2}{4 \pi} \bigg[ 1 + \frac{\tilde{g}^2 \mathrm{coth}\big( \tfrac{\beta \omega}{2} \big)  }{16 \pi^2 |\bfx_{\ssQ}|^2 \left[  (\omega/c_0)^2 + 1 \right]} \bigg] \\
\frac{\lambda_{\ssQ}^2\cD}{\omega} & \simeq & \frac{\lambda_{\ssQ}^2}{2\pi^2} \bigg[ \log\left( \frac{\omega}{\Lambda} \right) - \frac{\epsilon^2}{|\bfx_{\ssQ}|^2} \cdot \frac{1}{ (\omega/c_0)^2 + 1 } \cdot \log\left( \frac{\omega}{c_0} \right) \nn \\
& \ & \qquad \qquad \qquad \qquad + \frac{\tilde{g}^2}{16 \pi^2 |\bfx_{\ssQ}|^2} \cdot \frac{1}{(\omega/c_0)^2 + 1} \cdot \bigg\{  \mathrm{Re}\left[ \psi^{(0)}\left( i \frac{\beta \omega}{2\pi} \right) \right]  - \psi^{(0)}\left( \frac{\beta c_0}{2\pi} \right)- \frac{\pi}{\beta c_0} \bigg\} \bigg] \,. \nn 
\eea
Notice that these functions are nontrivial functions of the dimensionless variables $\beta \omega$,  $\beta c_0$ (and so also $\omega/ c_0$), whose absolute sizes are not determined solely using the conditions \pref{parameters2}. Table \ref{nonDegenTable1} explores the limiting form of these functions in various parametric regimes for which $\omega/c_0$ is small, $\cO(1)$ and large, with $\beta \omega$ either small or large. 


\begin{table}[h] 
\centering    
\centerline{\begin{tabular}{ c|c|c|c }
\multicolumn{1}{r}{}
& \multicolumn{1}{c}{$\underset{\ }{ \left| \dfrac{\lambda_{\ssQ}^2 \cC}{\omega} \right| \ll 1 }$}
& \multicolumn{1}{c}{$\left| \dfrac{\lambda_{\ssQ}^2 \cD}{\omega} \right| \ll 1$} \\
\cline{2-3}
$\beta \omega \ll \beta c_0 \ll 1$ & $\stackrel{\ }{ \underset{\ }{ \frac{ \lambda_{\ssQ }^2}{4\pi} \left| 1 + \frac{\tilde{g}^2}{16 \pi^2 |\bfx_{\ssQ}|^2} \cdot \frac{2}{\beta \omega}  \right| \ \ll \ 1 } }$ & $\stackrel{\ }{ \underset{\ }{ \frac{ \lambda_{\ssQ }^2}{2\pi^2} \left| \log\left( \frac{\omega}{\Lambda} \right) - \frac{\epsilon^2}{|\bfx_{\ssQ}|^2} \log\big( \frac{\omega}{c_0} \big) + \frac{\tilde{g}^2}{16 \pi^2 |\bfx_{\ssQ}|^2} \cdot \frac{\pi}{\beta c_0}   \right|  \ \ll \ 1 } }$ & \multirow{8}{*}[-0.4ex]{\rotatebox{90}{High Temp\ $(\beta \omega \ll 1)$}}  \\
 \cline{2-3}
$\beta \omega \ll 1 \ll \beta c_0$ & $\stackrel{\ }{ \underset{\ }{ \frac{ \lambda_{\ssQ }^2}{4\pi} \left|  1 + \frac{\tilde{g}^2}{16 \pi^2 |\bfx_{\ssQ}|^2} \cdot \frac{2}{\beta \omega}  \right| \ \ll \ 1 } }$ & $\stackrel{\ }{ \underset{\ }{ \frac{ \lambda_{\ssQ }^2}{2\pi^2} \left| \log\left( \frac{\omega}{\Lambda} \right) - \frac{\epsilon^2}{|\bfx_{\ssQ}|^2} \log\big( \frac{\omega}{c_0} \big) - \frac{\tilde{g}^2}{16 \pi^2 |\bfx_{\ssQ}|^2} \log\big( \frac{c_0 \beta}{2\pi} e^{\gamma} \big)  \right| \ \ll \ 1 } }$ \\
\cline{2-3}
$\beta \omega \simeq \beta c_0 \ll 1$ & $\stackrel{\ }{ \underset{\ }{ \frac{ \lambda_{\ssQ }^2}{4\pi} \left|  1 + \frac{\tilde{g}^2}{16 \pi^2 |\bfx_{\ssQ}|^2} \cdot \frac{1}{\beta \omega}  \right| \ \ll \ 1 } }$ & $\stackrel{\ }{ \underset{\ }{ \frac{ \lambda_{\ssQ }^2}{2\pi^2} \left| \log\left( \frac{\omega}{\Lambda} \right)  + \frac{\tilde{g}^2}{16 \pi^2 |\bfx_{\ssQ}|^2} \cdot \frac{\pi}{ 2 \beta \omega}   \right|  \ \ll \ 1 } }$ \\
\cline{2-3}
$\beta c_0 \ll \beta \omega \ll 1$ & $\stackrel{\ }{ \underset{\ }{ \frac{ \lambda_{\ssQ }^2}{4\pi} \left| 1 + \frac{\tilde{g}^2}{16 \pi^2 |\bfx_{\ssQ}|^2} \cdot \frac{2c_0^2}{\beta \omega^3}  \right| \ \ll \ 1 } }$ & $\stackrel{\ }{ \underset{\ }{ \frac{ \lambda_{\ssQ }^2}{2\pi^2} \left| \log\left( \frac{\omega}{\Lambda} \right)- \frac{\tilde{g}^2}{16 \pi^2 |\bfx_{\ssQ}|^2} \cdot \frac{c_0^2}{\omega^2} \cdot \log\big( \frac{c_0 \beta}{2\pi} e^{\gamma} \big) \right| \ \ll \ 1 } }$\\
\Xhline{3\arrayrulewidth}
$1 \ll \beta \omega \ll \beta c_0 $ & $\stackrel{\ }{ \underset{\ }{ \frac{ \lambda_{\ssQ }^2}{4\pi} \left| 1 + \frac{\tilde{g}^2}{16 \pi^2 |\bfx_{\ssQ}|^2}  \right| \ \ll \ 1 } }$ & $\stackrel{\ }{ \underset{\ }{ \frac{ \lambda_{\ssQ }^2}{2\pi^2} \left| \log\left( \frac{\omega}{\Lambda} \right) - \frac{\epsilon^2}{|\bfx_{\ssQ}|^2} \log\big( \frac{\omega}{c_0} \big) + \frac{\tilde{g}^2}{16 \pi^2 |\bfx_{\ssQ}|^2} \log\big( \frac{\omega}{c_0} \big) \right| \ \ll \ 1 } }$ & \multirow{8}{*}[-0.0ex]{\rotatebox{90}{Low Temp\ $(\beta \omega \gg 1)$}}  \\
 \cline{2-3}
$1 \ll \beta \omega \simeq \beta c_0$ & $\stackrel{\ }{ \underset{\ }{ \frac{ \lambda_{\ssQ }^2}{4\pi} \left| 1 + \frac{\tilde{g}^2}{16 \pi^2 |\bfx_{\ssQ}|^2} \cdot \frac{1}{2} \right| \ \ll \ 1 } }$ & $\stackrel{\ }{ \underset{\ }{ \frac{ \lambda_{\ssQ }^2}{2\pi^2} \left| \log\left( \frac{\omega}{\Lambda} \right)  + \frac{\tilde{g}^2}{16 \pi^2 |\bfx_{\ssQ}|^2} \cdot \frac{\pi^2}{ 3 \beta^2 \omega^2} \right| \ \ll \ 1 } }$ \\
\cline{2-3}
$1 \ll \beta c_0 \ll \beta \omega$ & $\stackrel{\ }{ \underset{\ }{ \frac{ \lambda_{\ssQ }^2}{4\pi} \left| 1 + \frac{\tilde{g}^2}{16 \pi^2 |\bfx_{\ssQ}|^2} \cdot \frac{c_0^2}{\omega^2} \right| \ \ll \ 1 } }$ & $\stackrel{\ }{ \underset{\ }{ \frac{ \lambda_{\ssQ }^2}{2\pi^2} \left| \log\left( \frac{\omega}{\Lambda} \right)- \frac{\tilde{g}^2}{16 \pi^2 |\bfx_{\ssQ}|^2} \cdot \frac{c_0^2}{\omega^2} \cdot \log\big( \frac{\omega}{c_0}\big) \right| \ \ll \ 1 } }$ \\
\cline{2-3}
$\beta c_0 \ll 1 \ll \beta \omega$ & $\stackrel{\ }{ \underset{\ }{ \frac{ \lambda_{\ssQ }^2}{4\pi} \left| 1 + \frac{\tilde{g}^2}{16 \pi^2 |\bfx_{\ssQ}|^2} \cdot \frac{c_0^2}{\omega^2}  \right| \ \ll \ 1 } }$ & $\stackrel{\ }{ \underset{\ }{ \frac{ \lambda_{\ssQ }^2}{2\pi^2} \left| \log\left( \frac{\omega}{\Lambda} \right)- \frac{\tilde{g}^2}{16 \pi^2 |\bfx_{\ssQ}|^2} \cdot \frac{c_0^2}{\omega^2} \cdot \big[ \frac{\pi}{\beta c_0} + \log\big( \frac{\beta \omega}{2\pi} e^{\gamma} \big) \big] \right| \ \ll \ 1 } }$ \\
 \cline{2-3}
\end{tabular} }
      \caption{The asymptotic forms for various relative sizes of $\beta \omega$ and $\beta c_0 $ the two quantities that must be small as in (\ref{nondegen}) to work with nondegenerate perturbation theory (see \cite{Kaplanek:2019dqu}) (these functions are explicitly written in \pref{NonDegenExplicit} in the parameter regime \pref{parameters} of interest).} \label{nonDegenTable1}
\end{table}

\vspace{2mm}

For each choice of parameter regime (or row in the Table) the nondegeneracy condition \pref{nondegen} requires the entries in each column to be much smaller than unity. Inspection of the table shows that this is often automatically satisfied by the condition $\lambda^2_\ssQ/4\pi \ll 1$ that is in any case required for use of perturbative methods.

\subsection{Markovian regime - part I}

Next we need to determine the regime of parameter space in which the Markovian approximation is valid. Recalling that (\ref{taylorMarkov}) keeps only the leading-order term of the Taylor series $\boldsymbol{\varrho}(t - s) \simeq \boldsymbol{\varrho}(t) - s \dot{\boldsymbol{\varrho}}(t) + \ldots$, we ask here when the first nominally sub-leading terms are small. 

As shown in detail in \cite{Kaplanek:2019dqu,Kaplanek:2020iay,Kaplanek:2019vzj} these sub-leading terms are small (and so the Markovian approximation applies) only when the following four conditions are satisfied:
\be \label{App:MarkovianConditions}
\left| \lambda_{\ssQ}^2 \frac{\exd \cC}{\exd \omega} \right| \ll 1 \ \ , \qquad \left| \lambda_{\ssQ}^2 \frac{\exd \cD}{\exd \omega} \right| \ll 1 \ \ , \qquad \left| \frac{\omega}{\cC} \cdot \frac{\exd \cC}{\exd \omega} \right| \ll 1 \quad \mathrm{and} \qquad \left| \frac{\omega}{\cC} \cdot  \frac{\exd \cD}{\exd \omega} \right| \ll 1 \ .
\ee
Explicit forms for the first two of these functions --- $\lambda_{\ssQ}^2 \frac{\mathrm{d} \cC}{\mathrm{d} \omega}$ and $\lambda_{\ssQ}^2 \frac{\mathrm{d} \cD}{\mathrm{d} \omega}$ --- are given for the parameter regime \pref{parameters} are given in \pref{Cprime} and \pref{Dprime}, and their approximate form for various choices for the sizes of $\omega/c_0$ and $\beta \omega$ are given in Table \ref{MarkovianTable1}. As this table shows, these quantities are again often automatically small in the perturbative regime, for which $\lambda_\ssQ^2/4\pi \ll 1$.

\begin{table}[h] 
\centering    
\centerline{\begin{tabular}{ c|c|c|c }
\multicolumn{1}{r}{}
& \multicolumn{1}{c}{$\underset{\ }{ \left|  \lambda_{\ssQ}^2 \dfrac{\exd \cC}{\exd \omega} \right| \ll 1 }$}
& \multicolumn{1}{c}{$\left| \lambda_{\ssQ}^2 \dfrac{\exd \cD}{\exd \omega} \right| \ll 1$} \\
\cline{2-3}
$\beta \omega \ll \beta c_0 \ll 1$ & $\stackrel{\ }{ \underset{\ }{ \frac{ \lambda_{\ssQ }^2}{4\pi} \left| 1 + \frac{\tilde{g}^2}{16\pi^2 |\bfx_{\ssQ}|^2} \big( \frac{\beta \omega}{3} - \frac{4\omega}{\beta c_0^2} \big) \right| \ \ll \ 1 } }$ & $\stackrel{\ }{ \underset{\ }{ \frac{ \lambda_{\ssQ }^2}{2\pi^2} \left| \log\big( \frac{\omega e^1}{\Lambda}  \big) - \frac{\epsilon^2 \log( {\omega e^1}/{c_0} )}{|\bfx_{\ssQ}|^2}  + \frac{\tilde{g}^2}{16\pi^2 |\bfx_{\ssQ}|^2} \cdot \frac{\pi}{\beta c_0} \right| \ \ll \ 1 } }$ & \multirow{8}{*}[-0.4ex]{\rotatebox{90}{High Temp\ $(\beta \omega \ll 1)$}}  \\
 \cline{2-3}
$\beta \omega \ll 1 \ll \beta c_0$ & $\stackrel{\ }{ \underset{\ }{ \frac{ \lambda_{\ssQ }^2}{4\pi} \left| 1 + \frac{\tilde{g}^2}{16\pi^2 |\bfx_{\ssQ}|^2} \big( \frac{\beta \omega}{3} - \frac{4\omega}{\beta c_0^2} \big) \right|  \ \ll \ 1 } }$ & $\stackrel{\ }{ \underset{\ }{ \frac{ \lambda_{\ssQ }^2}{2\pi^2} \left| \log\big( \frac{\omega e^1}{\Lambda} \big) - \frac{\epsilon^2 \log( {\omega e^1}/{c_0} )}{|\bfx_{\ssQ}|^2}  - \frac{\tilde{g}^2 \log( {c_0 \beta} \frac{e^{\gamma}}{2\pi} )}{16 \pi^2 |\bfx_{\ssQ}|^2}  \right| \ \ll \ 1 } }$ \\
\cline{2-3}
$\beta \omega \simeq \beta c_0 \ll 1$ & $\stackrel{\ }{ \underset{\ }{ \frac{ \lambda_{\ssQ }^2}{4\pi} \left| 1 - \frac{\tilde{g}^2}{16 \pi^2 |\bfx_{\ssQ}|^2} \cdot \frac{1}{\beta \omega}  \right| \ \ll \ 1 } }$ & $\stackrel{\ }{ \underset{\ }{ \frac{ \lambda_{\ssQ }^2}{2\pi^2} \left| \log\big( \frac{\omega e^1}{\Lambda}  \big) + \frac{\tilde{g}^2}{16 \pi^2 |\bfx_{\ssQ}|^2} \cdot \frac{\zeta(3) (\beta \omega)^2}{4\pi^2} \right| \ \ll \ 1 } }$ \\
\cline{2-3}
$\beta c_0 \ll \beta \omega \ll 1$ & $\stackrel{\ }{ \underset{\ }{ \frac{ \lambda_{\ssQ }^2}{4\pi} \left| 1 - \frac{\tilde{g}^2}{16 \pi^2 |\bfx_{\ssQ}|^2} \cdot \frac{4 c_0^2}{\beta \omega^3}  \right| \ \ll \ 1 } }$ & $\stackrel{\ }{ \underset{\ }{ \frac{ \lambda_{\ssQ }^2}{2\pi^2} \left| \log\big( \frac{\omega e^1 }{\Lambda} \big) - \frac{\tilde{g}^2}{16 \pi^2 |\bfx_{\ssQ}|^2} \cdot \frac{\pi c_0}{\beta \omega^2}  \right| \ \ll \ 1 } }$\\
\Xhline{3\arrayrulewidth}
$1 \ll \beta \omega \ll \beta c_0 $ & $\stackrel{\ }{ \underset{\ }{ \frac{ \lambda_{\ssQ }^2}{4\pi} \left| 1 + \frac{\tilde{g}^2}{16\pi^2 |\bfx_{\ssQ}|^2}\right| \ \ll \ 1 } }$ & $\stackrel{\ }{ \underset{\ }{ \frac{ \lambda_{\ssQ }^2}{2\pi^2} \left| \log\big( \frac{\omega e^1 }{\Lambda} \big) - \frac{\epsilon^2 \log( {\omega e^1}/{c_0} )}{|\bfx_{\ssQ}|^2}  + \frac{\tilde{g}^2 \log( {\omega e^1}/{c_0} )}{16 \pi^2 |\bfx_{\ssQ}|^2}  \right| \ \ll \ 1 } }$ & \multirow{8}{*}[-0.2ex]{\rotatebox{90}{Low Temp\ $(\beta \omega \gg 1)$}}  \\
 \cline{2-3}
$1 \ll \beta \omega \simeq \beta c_0$ & $\stackrel{\ }{ \underset{\ }{ \frac{ \lambda_{\ssQ }^2}{4\pi} \ \ll \ 1 } }$ & $\stackrel{\ }{ \underset{\ }{ \frac{ \lambda_{\ssQ }^2}{2\pi^2} \left| \log\big( \frac{\omega e^1}{\Lambda} \big) + \frac{\tilde{g}^2}{16 \pi^2 |\bfx_{\ssQ}|^2} \cdot \frac{1}{2} \right| \ \ll \ 1 } }$ \\
\cline{2-3}
$1 \ll \beta c_0 \ll \beta \omega$ & $\stackrel{\ }{ \underset{\ }{ \frac{ \lambda_{\ssQ }^2}{4\pi} \left| 1 - \frac{\tilde{g}^2}{16 \pi^2 |\bfx_{\ssQ}|^2} \cdot \frac{c_0^2}{\omega^2}  \right| \ \ll \ 1 } }$ & $\stackrel{\ }{ \underset{\ }{ \frac{ \lambda_{\ssQ }^2}{2\pi^2} \left| \log\big( \frac{\omega e^1}{\Lambda} \big) - \frac{\tilde{g}^2}{16 \pi^2 |\bfx_{\ssQ}|^2} \cdot \frac{c_0^2}{\omega^2} \log\big( \frac{\omega e^1 }{c_0} \big)  \right|  \ \ll \ 1 } }$ \\
\cline{2-3}
$\beta c_0 \ll 1 \ll \beta \omega$ & $\stackrel{\ }{ \underset{\ }{ \frac{ \lambda_{\ssQ }^2}{4\pi} \left| 1 - \frac{\tilde{g}^2}{16 \pi^2 |\bfx_{\ssQ}|^2} \cdot \frac{c_0^2}{\omega^2}  \right| \ \ll \ 1 } }$ & $\stackrel{\ }{ \underset{\ }{ \frac{ \lambda_{\ssQ }^2}{2\pi^2} \left| \log\big( \frac{\omega e^1}{\Lambda} \big) - \frac{\tilde{g}^2}{16 \pi^2 |\bfx_{\ssQ}|^2} \cdot \frac{\pi c_0}{\beta \omega^2} \right| \ \ll \ 1 } }$ \\
 \cline{2-3}
\end{tabular} }
      \caption{The asymptotic forms for the conditions $\lambda_{\ssQ}^2 \frac{\mathrm{d} \cC}{\mathrm{d} \omega} \ll 1$ and $\lambda_{\ssQ}^2 \frac{\mathrm{d} \cD}{\mathrm{d} \omega} \ll 1$ required in the Markovian limit (see \cite{Kaplanek:2019dqu}).} \label{MarkovianTable1}
\end{table}

Controlling the Markovian approximation is subtle, and conditions \pref{App:MarkovianConditions} are important when understanding why. For instance, the sharp-eyed reader may have noticed that formulae like \pref{12eqNZ4} appear to be missing terms that would naively be there if one keeps only the $s$-independent terms of the Taylor expansions for $\varrho_{ij}(t-s)$ in \pref{12eqNZ3}. The naive result would instead have looked like
\be
\frac{\pd \varrho_{12}(t)}{\pd t}   \simeq   -  \lambda_{\ssQ}^2 \Bigl[ \cC + i ( \omega_{\rm ct} +  \cD ) \Bigr] \varrho_{12}(t) \;  +  \;  \lambda_{\ssQ}^2 \, e^{+ 2 i \omega t } \, ( \cC - i \cD ) \, \varrho^{\ast}_{12}(t) \,,
\ee
with solution (in the non-degenerate limit of \pref{nondegen}, and after picking the counter-term $\omega_{\mathrm{ct}} = - \cD$)
\be \label{12Ddep}
\varrho_{12}(t) \ \simeq \ \bigg[ \varrho_{12}(t_0) + \varrho_{12}^{\ast}(t_0) \bigg( \frac{\lambda_{\ssQ}^2 \cD}{2\omega} + i \frac{\lambda_{\ssQ}^2 \cC}{2\omega} \bigg) (e^{2 i \omega t_0} - e^{2 i \omega t}) \bigg] e^{ - \lambda_{\ssQ}^2 \cC (t-t_0)}  \,. 
\ee
What would be bothersome about this result is that it explicitly depends on the divergent function $\cD$, but does not do so always together with the associated counterterm $\omega_{\rm ct}$. However, as we now show, this $\cD$-dependence is actually subdominant (and so can be dropped) in the regime for which the Markovian approximation applies.

As can be seen from Table \ref{nonDegenTable1} and \ref{MarkovianTable1}, $| \cD/\omega | \sim \big| \frac{\mathrm{d}}{\mathrm{d}\omega} \cD\big|$ in the various regimes of $\beta \omega$ and $\beta c_0$ considered, and this implies that 
\be
\left| \frac{\lambda_{\ssQ}^2 \cD}{\omega} \right| \ \simeq \ \left| \lambda_{\ssQ}^2 \frac{\mathrm{d} \cD}{\mathrm{d} \omega} \right| \ = \ \left| \frac{\lambda_{\ssQ}^2 \cC}{\omega} \right| \times \left| \frac{\omega}{\cC} \frac{\exd \cD}{\exd \omega} \right| \ \ll \ \left| \frac{\lambda_{\ssQ}^2 \cC}{\omega} \right| \,,
\ee
is {\it doubly} suppressed within the Markovian regime. Consequently $|{\lambda_{\ssQ}^2 \cD}/{\omega} | \ll | {\lambda_{\ssQ}^2 \cC}/{\omega} | \ll 1$ and so any $\cD$-dependence not accompanied by $\omega_{\rm ct}$ in the equations of motion drops out ({\it ie.} the factor $|{\lambda_{\ssQ}^2 \cD}/{\omega}|$ is negligibly small). In this case our earlier solution (\ref{12Ddep}) becomes instead
\be
\varrho_{12}(t) \ \simeq \ \bigg[ \varrho_{12}(t_0) + i \varrho_{12}^{\ast}(t_0) \frac{\lambda_{\ssQ}^2 \cC}{2\omega} (e^{2 i \omega t_0} - e^{ 2 i \omega t}) \bigg] e^{ - \lambda_{\ssQ}^2 \cC (t-t_0)}  \ , 
\ee
as quoted in (\ref{12sol1}).

\subsection{Markovian regime - part II}

Finally we extract the parameter information that is hidden in the conditions $\big| \frac{\omega}{\cC} \cdot \frac{\mathrm{d} }{\mathrm{d} \omega} \cC \big|  \ll 1$ and $ \big| \frac{\omega}{\cC} \cdot \frac{\mathrm{d}}{\mathrm{d} \omega} \cD \big| \ll 1$ of \pref{App:MarkovianConditions}. These particular conditions tend to be the most restrictive because in them all factors of $\lambda_\ssQ^2/4\pi$ cancel out, so the burden of making these terms small falls on the other parameters.

\begin{table}[h] 
\centering    
\centerline{\begin{tabular}{ c|c|c|c }
\multicolumn{1}{r}{}
& \multicolumn{1}{c}{$\underset{\ }{ \left| \dfrac{\omega}{\cC} \cdot \dfrac{\mathrm{d} \cC}{\mathrm{d} \omega} \right| \ll 1 }$}
& \multicolumn{1}{c}{$\left| \dfrac{\omega}{\cC} \cdot \dfrac{\mathrm{d} \cD}{\mathrm{d} \omega} \right| \ll 1$} \\
\cline{2-3}
$\beta \omega \ll \beta c_0 \ll 1$ & $\stackrel{\ }{ \underset{\ }{ \big| \frac{\omega^2 \beta^2}{6} - \frac{\omega^2}{c_0^2} \big| \ \ll \ 1 } }$ & $\stackrel{\ }{ \underset{\ }{ \frac{\omega}{c_0}  \ \ll \ 1 } }$ & \multirow{8}{*}[1.1ex]{\rotatebox{90}{High Temp $(\beta \omega \ll 1)$}}  \\
 \cline{2-3}
$\beta \omega \ll 1 \ll \beta c_0$ & $\stackrel{\ }{ \underset{\ }{ \big| \frac{\omega^2 \beta^2}{6} - \frac{\omega^2}{c_0^2} \big| \ \ll \ 1 } }$ & $\stackrel{\ }{ \underset{\ }{ \big| \frac{\beta \omega}{\pi} \log\big( \frac{\beta c_0}{2\pi} e^{\gamma} \big) \big| \ \ll \ 1 } }$ \\
\cline{2-3}
$\beta \omega \simeq \beta c_0 \ll 1$ & \cellcolor{pink} $\stackrel{\ }{ \underset{\ }{ 1 \ \ll \ 1 } }$ & $\stackrel{\ }{ \underset{\ }{ \frac{\pi^2 (\beta\omega)^2}{24} \ \ll \ 1 } }$ \\
\cline{2-3}
$\beta c_0 \ll \beta \omega \ll 1$ & $\cellcolor{pink} \stackrel{\ }{ \underset{\ }{\frac{\omega^2}{c_0^2} \ \ll \ 1 } }$ & $\stackrel{\ }{ \underset{\ }{ \frac{\zeta(3) (\beta \omega)^3}{16\pi} \ \ll \ 1 } }$ \\
\Xhline{3\arrayrulewidth}
$1 \ll \beta \omega \ll \beta c_0 $ & $\cellcolor{pink} \stackrel{\ }{ \underset{\ }{ 1 \ \ll \ 1 } }$ & \cellcolor{pink} $\stackrel{\ }{ \underset{\ }{ \frac{2}{\pi} \left| \log\big( \frac{\omega e^{1}}{c_0}  \big) \right|\ \ll \ 1 } }$ & \multirow{8}{*}[1.4ex]{\rotatebox{90}{Low Temp $(\beta \omega \gg 1)$}}  \\
 \cline{2-3}
$1 \ll \beta \omega \simeq \beta c_0$ & $\stackrel{\ }{ \underset{\ }{ 2 \beta \omega e^{- \beta \omega} \ \ll \ 1 } }$ & \cellcolor{pink} $\stackrel{\ }{ \underset{\ }{ \frac{\pi}{2} \ \ll \ 1 } }$ \\
\cline{2-3}
$1 \ll \beta c_0 \ll \beta \omega$ & $\cellcolor{pink} \stackrel{\ }{ \underset{\ }{ \frac{\omega^2}{c_0^2}  \ \ll \ 1 } }$ & \cellcolor{pink} $\stackrel{\ }{ \underset{\ }{ \frac{3\pi}{2} \log\big( \frac{\omega}{c_0} e^{-1/3} \big) \ \ll \ 1 } }$ \\
\cline{2-3}
$\beta c_0 \ll 1 \ll \beta \omega$ & $ \stackrel{\ }{ \underset{\ }{ \cellcolor{pink}  1 \ \ll \ 1  } }$ & \cellcolor{pink} $\stackrel{\ }{ \underset{\ }{ \frac{3\pi}{2} \log\big( \frac{\beta\omega}{2\pi} e^{\gamma - 1/3} \big) \ \ll \ 1  } }$ \\
 \cline{2-3}
\end{tabular} }
      \caption{The asymptotic forms for the conditions $\frac{\omega}{\cC} \cdot \frac{\mathrm{d} \cC}{\mathrm{d} \omega} \ll 1$ and $\frac{\omega}{\cC} \cdot \frac{\mathrm{d} \cD}{\mathrm{d} \omega} \ll 1$ (see the functions \pref{MarkovConditionsExplicitC} and \pref{MarkovConditionsExplicitD} above) required in the Markovian limit (see \cite{Kaplanek:2019dqu}). The cells coloured in pink belong to the regime in which the conditions are impossible to satisfy (notice only $\beta \omega \ll 1$ is possible here, and also we need $\omega \ll c_0$ as well). } \label{MarkovianTable2}
\end{table}

In the parameter regime \pref{parameters2} these particular functions take the following approximate form
\be \label{MarkovConditionsExplicitC}
\dfrac{\omega}{\cC} \cdot \dfrac{\mathrm{d} \cC}{\mathrm{d} \omega} \simeq  \frac{ \dfrac{  \mathrm{coth}\big( \tfrac{\beta \omega}{2} \big) - \frac{\beta \omega}{2} \mathrm{csch}^2 \big( \tfrac{\beta \omega}{2} \big)  }{ (\omega/c_0)^2 + 1 } - \dfrac{2 (\omega/c_0)^2 \coth\big( \tfrac{\beta \omega}{2} \big)}{\left[ (\omega/c_0)^2 +1 \right]} }{ \dfrac{\mathrm{coth}\big( \tfrac{\beta \omega}{2} \big)  }{ (\omega/c_0)^2 + 1 } }  \,,
\ee
and
\be \label{MarkovConditionsExplicitD}
\dfrac{\omega}{\cC} \cdot \dfrac{\mathrm{d} \cD}{\mathrm{d} \omega}  \simeq   \frac{\pi}{2} \cdot \frac{ \sfrac{1}{(\omega/c_0)^2 + 1} \cdot \left\{ \sfrac{\beta \omega}{2\pi}  \mathrm{Im}\left[ \psi^{(1)}\left( i \sfrac{\beta \omega}{2\pi} \right) \right] + \sfrac{3(\omega/c_0)^2 +1}{(\omega/c_0)^2 + 1} \left(  \mathrm{Re}\left[ \psi^{(0)}\left( i \sfrac{\beta \omega}{2\pi} \right) \right] - \psi^{(0)}\left( \sfrac{\beta c_0}{2\pi} \right)- \sfrac{\pi}{\beta c_0} \right) \right\} }{ \dfrac{\mathrm{coth}\big( \tfrac{\beta \omega}{2} \big)  }{ (\omega/c_0)^2 + 1 } } 
\ee
which both drop powers of ${16\pi^2 |\bfx_{\ssQ}|^2}/{\tilde{g}^2} \ll 1$.

The size of these functions for various choices of $\beta \omega$ and $\beta c_0$ are shown in Table \ref{MarkovianTable2}. In particular this Table shows that the Markovian approximation can only be attained in the high temperature limit where $\beta \omega \ll 1$ with $\omega \ll c_0$ (although $\beta c_0$ need not be either large or small).

\end{document}